\documentclass{article}
\usepackage{amsfonts}
\usepackage{amsmath}

\begin{document}

\begin{center}
{\LARGE Quantum integrability of the deformed elliptic Calogero--Moser
problem\vspace{0.7cm}}

{\large L. A. Khodarinova\vspace{0.3cm}}

\textit{Magnetic Resonance Centre, School of Physics and Astronomy,
University of Nottingham, Nottingham, England NG7 2RD, e-mail:
LarisaKhodarinova@hotmail.com}\vspace{1cm}
\end{center}

\textbf{Abstract:}\textit{\ }The integrability of the deformed quantum
elliptic Calogero-Moser problem introduced by Chalykh, Feigin and Veselov is
proven. Explicit recursive formulae for the integrals are found. For integer
values of the parameter this implies the algebraic integrability of the
systems.


\section{Introduction}

Deformed quantum Calogero--Moser (CM) systems were introduced by Chalykh,
Feigin and Veselov \cite{CFV1996, CFV1998}, who proved their integrability
in rational and trigonometric cases and conjectured that the same is true in
the elliptic case. The aim of this paper is to prove this conjecture.

Elliptic deformed CM system corresponds to the following Schr\"{o}dinger
operator 
\begin{equation}
L_{m}^{(n)}=\sum_{j=1}^{n}\frac{\hat{p}_{j}^{2}}{m_{j}}+2m(m+1)%
\sum_{j<k}^{n}m_{j}m_{k}\wp \left( x_{j}-x_{k}\right) ,  \label{operator}
\end{equation}%
where all but one \textquotedblleft masses\textquotedblright\ are equal, $%
m_{1}=m^{-1},\ m_{2}=\ldots =m_{n}=1,$ $m$ is a real parameter, $\hat{p}%
_{j}=i\frac{\partial }{\partial x_{j}},$ $j=1,\ldots ,n,$ and $\wp $ is the
classical Weierstrass elliptic function. The case when $m$ is integer is a
special one: in that case a stronger version of integrability (the so-called
algebraic integrability) was conjectured \cite{CFV1998}. The first results
in this direction were found in~\cite{KP2001}, where it was proven in the
simplest non-trivial case $n=3,m=2.$

The main result of the present paper is an explicit recursive formula for
the quantum integrals of the elliptic deformed CM system. This proves
integrability of the system for all $n$ and $m$ and due to a general recent
result by Chalykh, Etingof and Oblomkov \cite{CEO2003} this also implies the
algebraic integrability for integer values of the parameter~$m.$

As a by-product we have also new formulae for the integrals of the usual
quantum elliptic CM problem, which was the subject of many investigations
since 1970s (see in particular \cite{CMR1975}, \cite{SK1975}, \cite{OP1977,
OP1983}, \cite{OOS1994, OS1995, O1998}). We will be using some technical
tricks from these important papers. In trigonometric and rational limits we
have the formulae for the quantum integrals of the corresponding deformed CM
systems which are also seem to be new.

\section{Preliminaries and main formulae}

Quantum Hamiltonian of the deformed elliptic CM problem has the following
form 
\begin{equation}
H=-\left( m\partial _{1}^{2}+\partial _{2}^{2}+\ldots +\partial
_{n-1}^{2}+\partial _{n}^{2}\right) +2\left( m+1\right) \sum_{k=2}^{n}\wp
_{1k}+2m\left( m+1\right) \sum_{2\leq j<k\leq n}\wp _{jk},  \label{H}
\end{equation}%
where $\partial _{i}=\frac{\partial }{\partial x_{i}},$ $\wp _{jk}=\wp
\left( x_{j}-x_{k}\right) .$ Here $\wp $ is the classical Weierstrass
elliptic function~\cite{WW1927}, which can be determined by the differential
equation%
\begin{equation}
\left( \wp ^{\prime }\left( z\right) \right) ^{2}=4\left( \wp \left(
z\right) -e_{1}\right) \left( \wp \left( z\right) -e_{2}\right) \left( \wp
\left( z\right) -e_{3}\right) =4\wp ^{3}\left( z\right) -g_{2}\wp \left(
z\right) -g_{3}.  \label{differ_equation_P}
\end{equation}%
The Laurent expansion of $\wp $ at the origin is of the following form~\cite%
{WW1927}%
\begin{equation}
\wp \left( z\right) =z^{-2}+\sum_{k=1}^{\infty }\gamma _{2k}z^{2k},
\label{expansion_of_P}
\end{equation}%
where%
\begin{equation*}
\gamma _{2}=\frac{1}{20}g_{2},\quad \gamma _{4}=\frac{1}{28}g_{3}.
\end{equation*}%
The coefficients $\gamma _{2k}$ are related to the so-called \textit{%
Bernoulli-Hurwitz numbers} $BH\left( k\right) $ \cite{Katz}%
\begin{equation*}
\gamma _{2k}=\frac{1}{\left( 2k\right) !}\frac{BH\left( 2k+2\right) }{\left(
2k+2\right) }.
\end{equation*}%
There is a recursive formula which allows one to obtain the coefficients $%
\gamma _{2k+2}$ from the coefficients of the lower order:%
\begin{equation}
\gamma _{2k+2}=\frac{3}{\left( k-1\right) \left( 2k+5\right) }%
\sum_{j=1}^{k-1}\gamma _{2j}\gamma _{2k-2j},\quad k=2,3,\ldots .
\label{gammas_relationship}
\end{equation}%
The relationship~(\ref{gammas_relationship}) is easy to verify. One needs to
differentiate~(\ref{differ_equation_P}) to obtain%
\begin{equation*}
\wp ^{\prime \prime }\left( z\right) =6\wp ^{2}\left( z\right) -\frac{1}{2}%
g_{2},
\end{equation*}%
use the expansion~(\ref{expansion_of_P}) and collect terms at the
appropriate degrees of $z.$

To construct the integrals of the operator (\ref{H}) we will follow the idea
going back to \cite{SK1975}, \cite{OP1977}. Namely, the integrals are
constructed from the highest one by successive commutators with some
function which in our case is $m^{-1}x_{1}+x_{2}+\ldots +x_{n}.$ This
highest integral is of order $n$ and will be denoted below as $I.$ The rest
of this section is to explain the main ingredients of the formula for $I.$

Let us introduce the following differential operators $\mathfrak{D}^{k}$ in $%
\mathfrak{\partial }_{1}$ with constant coefficients%
\begin{gather}
\mathfrak{D}^{1}=\mathfrak{\partial }_{1},\quad \mathfrak{D}^{2}=\frac{%
\left( 1-m\right) }{2!}\mathfrak{\partial }_{1}^{2},\quad \mathfrak{D}^{3}=%
\frac{\left( 1-m\right) \left( 1-2m\right) }{3!}\mathfrak{\partial }%
_{1}^{3},\quad \mathfrak{D}^{4}=\frac{\left( 1-m\right) \left( 1-2m\right)
\left( 1-3m\right) }{4!}\mathfrak{\partial }_{1}^{4},  \notag \\
\mathfrak{D}^{k}=p_{0,k}\mathfrak{\partial }_{1}^{k}+\sum_{i=2}^{\left[ 
\frac{k}{2}\right] }p_{2i,k}\mathfrak{\partial }_{1}^{k-2i}.
\label{D_formulae}
\end{gather}%
The constants $p_{0,k}$ $,$ $k=2,3,\ldots ,$ are given by%
\begin{equation}
\begin{array}{c}
p_{0,k}=\frac{1}{k!}\prod_{l=1}^{k-1}\left( 1-lm\right) ,\quad k=2,3,\ldots ,%
\end{array}
\label{p_0,k}
\end{equation}%
and constants $p_{2i,k},$ $i=2,3,\ldots ,$ $k=2i,2i+1,\ldots ,$ are
determined by the following recursive relationship%
\begin{eqnarray}
&&%
\begin{array}{c}
p_{2i,2i}=0,\quad i=2,3,\ldots ,%
\end{array}
\label{p_2i,2i} \\
&&%
\begin{array}{l}
p_{2i,k}=\frac{\left( 1-m\left( k-2i-1\right) \right) }{k-2i}%
p_{2i,k-1}-\left( 1+m\right) \sum\limits_{\substack{ j=1  \\ j\neq 2}}%
^{i-1}\left( 2i-2j\right) !C_{k-2j+1}^{k-2i}\gamma _{2i-2j}p_{2j-2,k-1}, \\ 
i=2,3,\ldots ,\quad k=2i+1,2i+2,\ldots%
\end{array}
\label{p_2i,k}
\end{eqnarray}%
and $C_{n}^{k}=\frac{n!}{k!\left( n-k\right) !}$\ is a binomial coefficient.
To illustrate the formula (\ref{p_2i,k}) let us write explicitly the
formulae for the first few values of $i:$\ 
\begin{eqnarray*}
&&%
\begin{array}{c}
p_{4,k}=\frac{\left( 1-m\left( k-5\right) \right) }{k-4}p_{4,k-1}-\left(
1+m\right) \frac{\prod_{l=1}^{3}\left( k-l\right) }{3}\gamma
_{2}p_{0,k-1},\qquad k=5,6,\ldots ,%
\end{array}
\\
&&%
\begin{array}{c}
p_{6,k}=\frac{\left( 1-m\left( k-7\right) \right) }{k-6}p_{6,k-1}-\left(
1+m\right) \frac{\prod_{l=1}^{5}\left( k-l\right) }{5}\gamma
_{4}p_{0,k-1},\qquad k=7,8,\ldots ,%
\end{array}
\\
&&%
\begin{array}{c}
p_{8,k}=\frac{\left( 1-m\left( k-9\right) \right) }{k-8}p_{8,k-1}-\left(
1+m\right) \left( \frac{\prod_{l=1}^{7}\left( k-l\right) }{7}\gamma
_{6}p_{0,k-1}-\frac{\prod_{l=5}^{7}}{3}\gamma _{2}p_{4,k-1}\right) ,\qquad
k=9,10,\ldots .%
\end{array}%
\end{eqnarray*}%
It is interesting to note that for the special values of parameter $m=\frac{1%
}{l},$\ where $l$\ is a positive integer number, most of the constants (\ref%
{p_0,k}) and (\ref{p_2i,k}) are zero. For example, if $m=\frac{1}{2}$\ only $%
p_{0,2}$\ is non zero, if $m=\frac{1}{3}$\ then only $p_{0,2}$\ and $p_{0,3}$%
\ are non zero, if $m=\frac{1}{4}$\ then only $p_{0,2},$\ $p_{0,3},$\ $%
p_{0,4}$\ and $p_{4,5},$\ $p_{4,6}=p_{0,2}p_{4,5},$ $p_{4,7}=p_{0,3}p_{4,5},$
$p_{4,8}=p_{0,4}p_{4,5}$\ are non zero and so on.

Let us introduce the following notations: 
\begin{equation*}
\begin{array}{l}
\varsigma _{j}=\left( m+1\right) \varsigma \left( x_{1}-x_{j}\right) ,\qquad
2\leq j\leq n, \\ 
u_{1j}=\left( m+1\right) \wp \left( x_{1}-x_{j}\right) ,\qquad 2\leq j\leq n,
\\ 
u_{kl}=m\left( m+1\right) \wp \left( x_{k}-x_{l}\right) ,\qquad 2\leq
k<l\leq n,%
\end{array}%
\end{equation*}%
where $\varsigma $ is the standard elliptic $\varsigma $-function: $\frac{%
d\varsigma \left( z\right) }{dz}=-\wp \left( z\right) .$

We will need also to consider all the subsystems of the deformed CM system.
Let $S$ be a subset of the set $\left\{ 1,2,\ldots ,n\right\} $ and $\sigma
=\left\{ j_{1},j_{2},\ldots ,j_{t}\right\} ,$ $j_{1}<j_{2}<\ldots <j_{t},$\
be a subset of $t$ indices chosen from $S.$ The set of all different subsets 
$\sigma $ of size $t$ of the set $S$ will be denoted by 
\begin{equation*}
\mathfrak{S}\left( S;t\right) =\left\{ \sigma =\left\{ j_{1},j_{2},\ldots
,j_{t}\right\} :j_{1}<j_{2}<\ldots <j_{t},\;j_{k}\in S\right\} .
\end{equation*}%
If $\sigma \in \mathfrak{S}\left( S;t\right) $ define the set $S\backslash
\sigma =\left\{ j:j\in S\text{ and }j\notin \sigma \right\} .$ If $S=\left\{
1,2,\ldots ,n\right\} $ we will use short notation $\mathfrak{S}\left(
t\right) =\mathfrak{S}\left( \left\{ 1,2,\ldots ,n\right\} ;t\right) $ and $%
\hat{\sigma}=\left\{ 1,2,\ldots ,n\right\} \backslash \sigma .$ If $\sigma $%
\ contains only one element $\sigma =\left\{ k\right\} $ the brackets will
be omitted: $k$ will denote a set which contains one element $\left\{
k\right\} ,$\ $S\backslash k=S\backslash \left\{ k\right\} $ and $\hat{k}%
=\left\{ 1,2,\ldots ,n\right\} \backslash \left\{ k\right\} .$ We will also
use notation $\hat{\sigma}_{1}\hat{\sigma}_{2}$ to denote the intersection
of the subsets $\hat{\sigma}_{1}$ and $\hat{\sigma}_{2}:$ $\hat{\sigma}_{1}%
\hat{\sigma}_{2}=\hat{\sigma}_{1}\cap \hat{\sigma}_{2}.$

We use the notation $ad_{\varsigma _{\sigma }}^{t}\left( \mathfrak{D}%
^{k}\right) $ to denote the repeated commutator%
\begin{equation}
ad_{\varsigma _{\sigma }}^{t}\left( \mathfrak{D}^{k}\right) =\left[
\varsigma _{i_{t}},\ldots \left[ \varsigma _{i_{2}},\left[ \varsigma
_{i_{1}},\mathfrak{D}^{k}\right] \right] \right] ,\qquad \sigma =\left\{
j_{1},j_{2},\ldots ,j_{t}\right\} .  \label{add_with_dzeta}
\end{equation}%
Note that the order in which $\varsigma _{i_{k}}$ are used is not important
because of the form of the operator $\mathfrak{D}^{k}.$ Define 
\begin{equation*}
\Theta =\mathfrak{D}^{n}+\sum_{t=1}^{\left[ \frac{n}{2}\right] }\sum_{\sigma
\in \mathfrak{S}\left( \hat{1};t\right) }ad_{\varsigma _{\sigma }}^{t}\left( 
\mathfrak{D}^{n-t}\right) =\sum_{t=0}^{\left[ \frac{n}{2}\right]
}\sum_{\sigma \in \mathfrak{S}\left( \hat{1};t\right) }ad_{\varsigma
_{\sigma }}^{t}\left( \mathfrak{D}^{n-t}\right) .
\end{equation*}%
Let the set $S$ consist of $k$ elements and contain $1$, then define the
operator $\Theta _{S}$ as%
\begin{equation}
\Theta _{S}=\mathfrak{D}^{k}+\sum_{t=1}^{\left[ \frac{k}{2}\right]
}\sum_{\sigma \in \mathfrak{S}\left( S\backslash 1;t\right) }ad_{\varsigma
_{\sigma }}^{t}\left( \mathfrak{D}^{k-t}\right) =\sum_{t=0}^{\left[ \frac{k}{%
2}\right] }\sum_{\sigma \in \mathfrak{S}\left( S\backslash 1;t\right)
}ad_{\varsigma _{\sigma }}^{t}\left( \mathfrak{D}^{k-t}\right) .
\label{Theta}
\end{equation}%
We also use the notation $\partial _{S}$ to denote the product $\partial
_{S}=\prod\limits_{j\in S}\partial _{j}.$ If $1\in S$ then $\Delta _{S}$
denotes the sum $\Delta _{S}=m\partial _{1}^{2}+\sum_{j\in S\backslash
1}\partial _{j}^{2}.$ If $1\notin S$ then $\Delta _{S}=\sum_{j\in S}\partial
_{j}^{2}.$ By $I_{S}$ we will mean the corresponding quantum integral of the
system with the Hamiltonian%
\begin{equation*}
H_{S}=-\Delta _{S}+2\sum\limits_{\substack{ j<k  \\ j,k\in S}}u_{jk}.
\end{equation*}

Now we are ready to give the formula for the integrals:%
\begin{equation}
I=\sum_{t=1}^{n-2}\left( -1\right) ^{t+1}\sum_{\sigma \in \mathfrak{S}\left(
t\right) }I_{\hat{\sigma}}\partial _{\sigma }+\left( -1\right) ^{n}\left(
n-1\right) \partial _{1}\ldots \partial _{n}+X,  \label{I}
\end{equation}%
where%
\begin{equation}
X=\Theta +\sum\limits_{t=1}^{\left[ \frac{n-2}{2}\right] }\sum\limits_{%
\sigma \in \mathfrak{S}\left( \hat{1};2t\right) }X_{\sigma }\Theta _{\hat{%
\sigma}}  \label{X}
\end{equation}%
and $X_{\sigma }$ are related to a non deformed CM subsystem and\ are
determined by the recurrent formulae 
\begin{equation}
\begin{array}{c}
X_{\hat{1}}=\sum_{j=2}^{n-1}X_{\left\{ j,n\right\} }X_{\hat{1}\backslash
\left\{ j,n\right\} },\text{ if }n=2p-1;\text{ }X_{\hat{1}}=0,\text{ if }n=2p%
\text{ and }X_{\left\{ 2,3\right\} }=u_{23}.%
\end{array}
\label{X_no_def_formula}
\end{equation}

\textbf{Theorem 1.} \textit{The operator $I$ defined by (\ref{I}) and (\ref%
{X}) commutes with the deformed elliptic CM operator~$H$.}

\textbf{Remark.} The formula (\ref{I}) is valid in the non deformed case
also: the operator%
\begin{equation*}
I_{\hat{1}}=\sum_{t=1}^{n-3}\left( -1\right) ^{t+1}\sum_{\sigma \in 
\mathfrak{S}\left( \hat{1},t\right) }I_{\hat{\sigma}}\partial _{\sigma
}+\left( -1\right) ^{n}\left( n-2\right) \partial _{2}\ldots \partial
_{n}+X_{\hat{1}}
\end{equation*}%
commutes with the operator 
\begin{equation*}
H_{\hat{1}}=-\left( \partial _{2}^{2}+\ldots +\partial _{n-1}^{2}+\partial
_{n}^{2}\right) +2m\left( m+1\right) \sum_{2\leq j<k\leq n}\wp _{jk},
\end{equation*}%
which is the usual $n-1$ particle elliptic CM operator.

\textbf{Idea of the proof of Theorem 1. }The proof will be done by
induction. The main idea behind formula (\ref{I}) consists in the
observation that one can use the commutativity of $I_{\hat{\sigma}}\partial
_{\sigma }$ with $H_{\hat{\sigma}}$ to simplify the commutator $\left[ I,H%
\right] $ to the expression%
\begin{equation*}
\left[ I,H\right] =\left[ X,H\right] +\sum\nolimits_{j=1}^{n}\left[ X_{\hat{%
\jmath}}\partial _{j},2\sum\nolimits_{\substack{ l=1  \\ l\neq j}}^{n}u_{jl}%
\right] +\sum\nolimits_{1\leq k<l\leq n}\left[ X_{\hat{k}\hat{l}}\partial
_{k}\partial _{l},2u_{kl}\right] ,
\end{equation*}%
where terms $X,$ $X_{\hat{\jmath}}$ and $X_{\hat{k}\hat{l}}$ depend only on $%
\partial _{1}$ and $u_{sr},$ $1\leq s<r\leq n.$ This is shown in Lemma 1 in
section 6. At the next step we notice that if $X$ is given by (\ref{X}) the
commutator $\left[ I,H\right] $ can be simplified further to the following
expression%
\begin{equation*}
\left[ I,H\right] =\left[ \Theta ,H\right] +\sum\limits_{j=2}^{n}\left[
\Theta _{\hat{\jmath}}\partial _{j},2\sum\limits_{\substack{ l=1  \\ l\neq j
}}^{n}u_{jl}\right] +\sum\limits_{2\leq k<l\leq n}\left[ \Theta _{\hat{k}%
\hat{l}},2\left( u_{1k}+u_{1l}\right) \right] X_{\left\{ k,l\right\} }
\end{equation*}%
where $\Theta ,$ $\Theta _{\hat{\jmath}}$ and $\Theta _{\hat{k}\hat{l}}$
depend only on $\partial _{1}$ and $u_{1s},$ $2\leq s\leq n.$ From this we
can deduce that 
\begin{equation*}
\frac{\partial \Theta }{\partial x_{k}}=\left[ u_{1k},\Theta _{\hat{k}}%
\right]
\end{equation*}%
and, therefore, it seems natural to use the operators $ad_{\varsigma
_{\sigma }}^{t}$ to construct $\Theta .$ At this stage the only freedom left
is in choosing operators $\mathfrak{D}^{k}$ which must be the operators in $%
\partial _{1}$ with constant coefficients. To ensure that $\left[ I,H\right]
=0$ these operators must satisfy the relation%
\begin{equation*}
\begin{array}{l}
\left[ \mathfrak{D}^{n},\wp \left( x_{1}-x_{i}\right) \right] +\left(
1+m\right) \left[ \left[ \varsigma \left( x_{1}-x_{i}\right) ,\mathfrak{D}%
^{n-1}\right] ,\wp \left( x_{1}-x_{i}\right) \right] \\ 
\qquad +m\left[ \mathfrak{D}^{n-1},\wp \left( x_{1}-x_{i}\right) \right]
\partial _{1}-\frac{\left( 1+m\right) }{2}\wp ^{\prime }\left(
x_{1}-x_{i}\right) \mathfrak{D}^{n-1}-\frac{\left( 1-m\right) }{2}\mathfrak{D%
}^{n-1}\wp ^{\prime }\left( x_{1}-x_{i}\right) =0.%
\end{array}%
\end{equation*}%
which is equivalent to a large set of identities. It is remarkable that the
constants in $\mathfrak{D}^{k}$ can be chosen in such a way that all these
identities is satisfied (Lemma~3 of Section~6). The choice of the constants
is related to the Bernoulli-Hurwitz numbers and is described above. The
complete proof of the theorem is quite technical and is given in a separate
section.

\section{Examples: formulae for two, three and four particles.}

To illuminate our formulae let us consider more explicitly the case of small 
$n.$

\subsection{Two-particle case.}

In that case we have the operator 
\begin{equation*}
H=-m\partial _{1}^{2}-\partial _{2}^{2}+2\left( m+1\right) \wp
_{12}=-m\partial _{1}^{2}-\partial _{2}^{2}+2u_{12}
\end{equation*}%
which is trivially integrable since the operator $\partial _{1}+\partial
_{2} $ obviously commutes with it. This system gives the formula for
operator $I_{\left\{ 1,2\right\} }$ which starts the recursive construction
of the integral $I$ (\ref{I}). We have%
\begin{equation*}
\begin{array}{c}
I=I_{\left\{ 1,2\right\} }=\frac{1}{2}\left( H+\left( \partial _{1}+\partial
_{2}\right) ^{2}\right) =\partial _{1}\partial _{2}+\frac{\left( 1-m\right) 
}{2}\partial _{1}^{2}+\left( m+1\right) \wp _{12}=\partial _{1}\partial _{2}+%
\mathfrak{D}^{2}+u_{12}=\partial _{1}\partial _{2}+\Theta ,%
\end{array}%
\end{equation*}%
\begin{equation*}
X=X_{\left\{ 1,2\right\} }=\Theta _{\left\{ 1,2\right\} }=\mathfrak{D}%
^{2}+u_{12}=\mathfrak{D}^{2}+\left[ \varsigma _{2},\mathfrak{D}\right] .
\end{equation*}

\subsection{Three-particle case.}

The integrals for the problem of three particles have been found in~\cite%
{K1998}. The operator of the third order has the form 
\begin{eqnarray*}
I &=&%
\begin{array}{c}
\partial _{1}\partial _{2}\partial _{3}+\frac{\left( 1-m\right) }{2}%
(\partial _{2}+\partial _{3})\partial _{1}^{2}+\frac{\left( 1-m\right)
\left( 1-2m\right) }{3!}\partial _{1}^{3}+(m+1)\left( m\wp _{23}\partial
_{1}+\wp _{13}\partial _{2}+\wp _{12}\partial _{3}\right)%
\end{array}
\\
&&%
\begin{array}{c}
+\frac{\left( 1-m\right) }{2}(m+1)\left( \left( \wp _{12}+\wp _{13}\right)
\partial _{1}+\partial _{1}\left( \wp _{12}+\wp _{13}\right) \right)%
\end{array}
\\
&=&%
\begin{array}{c}
\partial _{1}\partial _{2}\partial _{3}+p_{0,2}(\partial _{2}+\partial
_{3})\partial _{1}^{2}+u_{23}\partial _{1}+u_{13}\partial
_{2}+u_{12}\partial _{3}+\mathfrak{D}^{3}+\left[ \varsigma _{2},\mathfrak{D}%
^{2}\right] +\left[ \varsigma _{3},\mathfrak{D}^{2}\right] ,%
\end{array}%
\end{eqnarray*}%
Operator $I$ can be rewritten as%
\begin{equation*}
I=I_{\left\{ 2,3\right\} }\partial _{1}+I_{\left\{ 1,3\right\} }\partial
_{2}+I_{\left\{ 1,2\right\} }\partial _{3}-2\partial _{1}\partial
_{2}\partial _{3}+X,
\end{equation*}%
where%
\begin{equation*}
X=\Theta =\mathfrak{D}^{3}+\left[ \varsigma _{2},\mathfrak{D}^{2}\right] +%
\left[ \varsigma _{3},\mathfrak{D}^{2}\right] ,\text{ and }I_{\left\{
2,3\right\} }=\partial _{2}\partial _{3}+u_{23}.
\end{equation*}

\subsection{Four-particle case.}

One can check by direct calculation that the operator 
\begin{eqnarray*}
I &=&%
\begin{array}{c}
\partial _{1}\partial _{2}\partial _{3}\partial _{4}+\frac{\left( 1-m\right) 
}{2}\left( \partial _{2}\partial _{3}+\partial _{2}\partial _{4}+\partial
_{3}\partial _{4}\right) \partial _{1}^{2}%
\end{array}
\\
&&%
\begin{array}{c}
+\frac{\left( 1-m\right) \left( 1-2m\right) }{3!}\left( \partial
_{2}+\partial _{3}+\partial _{4}\right) \partial _{1}^{3}+\frac{\left(
1-m\right) \left( 1-2m\right) \left( 1-3m\right) }{4!}\partial _{1}^{4}%
\end{array}
\\
&&%
\begin{array}{c}
+m(m+1)\left( \wp _{34}\partial _{1}\partial _{2}+\wp _{24}\partial
_{1}\partial _{3}+\wp _{23}\partial _{1}\partial _{4}\right) +(m+1)\left(
\wp _{14}\partial _{2}\partial _{3}+\wp _{13}\partial _{2}\partial _{4}+\wp
_{12}\partial _{3}\partial _{4}\right)%
\end{array}
\\
&&%
\begin{array}{c}
+\frac{\left( 1-m\right) }{2}\left( m+1\right) \partial _{1}\left( \left(
\wp _{13}+\wp _{14}\right) \partial _{2}+\left( \wp _{12}+\wp _{14}\right)
\partial _{3}+\left( \wp _{12}+\wp _{13}\right) \partial _{4}\right)%
\end{array}
\\
&&%
\begin{array}{c}
+\frac{\left( 1-m\right) }{2}\left( m+1\right) \left( \left( \wp _{13}+\wp
_{14}\right) \partial _{2}+\left( \wp _{12}+\wp _{14}\right) \partial
_{3}+\left( \wp _{12}+\wp _{13}\right) \partial _{4}\right) \partial _{1}%
\end{array}
\\
&&%
\begin{array}{c}
+\frac{\left( 1-m\right) \left( 1-2m\right) }{3!}\left( m+1\right) \left(
\left( \wp _{12}+\wp _{13}+\wp _{14}\right) \partial _{1}^{2}+\partial
_{1}\left( \wp _{12}+\wp _{13}+\wp _{14}\right) \partial _{1}+\partial
_{1}^{2}\left( \wp _{12}+\wp _{13}+\wp _{14}\right) \right)%
\end{array}
\\
&&%
\begin{array}{c}
+\frac{\left( 1-m\right) }{2}m(m+1)\left( \wp _{34}+\wp _{24}+\wp
_{23}\right) \partial _{1}^{2}%
\end{array}
\\
&&%
\begin{array}{c}
+\left( 1-m\right) \left( m+1\right) ^{2}(\wp _{14}\wp _{24}+\wp _{14}\wp
_{34}+\wp _{24}\wp _{34})+m\left( m+1\right) ^{2}\left( \wp _{12}\wp
_{34}+\wp _{13}\wp _{24}+\wp _{14}\wp _{23}\right)%
\end{array}%
\end{eqnarray*}%
commutes with $H.$ The operator $I$ can be rewritten in the following
recursive form%
\begin{eqnarray}
I &=&3\partial _{1}\partial _{2}\partial _{3}\partial _{4}+X+I_{\left\{
2,3,4\right\} }\partial _{1}+I_{\left\{ 1,3,4\right\} }\partial
_{2}+I_{\left\{ 1,2,4\right\} }\partial _{3}+I_{\left\{ 1,2,3\right\}
}\partial _{4}  \label{I_4} \\
&&-I_{\left\{ 3,4\right\} }\partial _{1}\partial _{2}-I_{\left\{ 2,4\right\}
}\partial _{1}\partial _{3}-I_{\left\{ 2,3\right\} }\partial _{1}\partial
_{4}-I_{\left\{ 1,4\right\} }\partial _{2}\partial _{3}-I_{\left\{
1,3\right\} }\partial _{2}\partial _{4}-I_{\left\{ 1,2\right\} }\partial
_{3}\partial _{4},  \notag
\end{eqnarray}%
where%
\begin{eqnarray*}
X &=&\Theta +X_{\left\{ 2,3\right\} }\Theta _{\left\{ 1,4\right\}
}+X_{\left\{ 2,4\right\} }\Theta _{\left\{ 1,3\right\} }+X_{\left\{
3,4\right\} }\Theta _{\left\{ 1,2\right\} } \\
&=&\mathfrak{D}^{4}+\left[ \varsigma _{2},\mathfrak{D}^{3}\right] +\left[
\varsigma _{3},\mathfrak{D}^{3}\right] +\left[ \varsigma _{4},\mathfrak{D}%
^{3}\right] +\left[ \varsigma _{2},\left[ \varsigma _{3},\mathfrak{D}^{2}%
\right] \right] +\left[ \varsigma _{2},\left[ \varsigma _{4},\mathfrak{D}^{2}%
\right] \right] +\left[ \varsigma _{3},\left[ \varsigma _{4},\mathfrak{D}^{2}%
\right] \right] \\
&&+u_{23}\left( \mathfrak{D}^{2}+\left[ \varsigma _{4},\mathfrak{D}\right]
\right) +u_{24}\left( \mathfrak{D}^{2}+\left[ \varsigma _{3},\mathfrak{D}%
\right] \right) +u_{34}\left( \mathfrak{D}^{2}+\left[ \varsigma _{2},%
\mathfrak{D}\right] \right) .
\end{eqnarray*}%
and%
\begin{equation*}
I_{\left\{ 2,3,4\right\} }=\partial _{1}\partial _{2}\partial
_{3}+u_{23}\partial _{1}+u_{13}\partial _{2}+u_{12}\partial _{3}.
\end{equation*}

\section{Integrability of the deformed elliptic quantum CM problem.}

Let us introduce the function $\theta =m^{-1}x_{1}+x_{2}+\ldots +x_{n}$ and
consider the corresponding "ad"-operation: 
\begin{equation*}
ad_{\theta }\left( L\right) =\left[ \theta ,L\right] .
\end{equation*}%
Following to the procedure known for the usual CM system (see, for example, 
\cite{OP1983}) consider the operators $L_{k}=ad_{\theta }^{k}\left( I\right)
,$ $k=0,1,\ldots ,n-1.$

\textbf{Theorem 2.} \textit{Operators $L_{k}=ad_{\theta }^{k}\left( I\right)
,$ }$k=0,1,\ldots ,n-1,$\textit{\ where $I$ are given by (\ref{I}) commute
with each other and with operator $H.$}

\textbf{Proof.} The proof is similar to the non-deformed case \cite{OP1983,
O1998}.

Let us first prove that $L_{k}$ commute with $H$ and with $%
\sum_{j=1}^{n}\partial _{j}.$ Proof is by induction in $k$. For $k=0$ it
follows from the Theorem 1. Let us assume that this is true for $k=i$, then
for $k=i+1$ we have by the Jacobi identity 
\begin{equation*}
\left[ H,L_{i+1}\right] =\left[ H,\left[ \theta ,L_{i}\right] \right] =\left[
\left[ H,\theta \right] ,L_{i}\right] +\left[ \left[ L_{i},H\right] ,\theta %
\right] =2\left[ \sum\nolimits_{j=1}^{n}\partial _{j},L_{i}\right] +\left[
0,\theta \right] =0
\end{equation*}%
and%
\begin{eqnarray*}
\left[ \sum\nolimits_{j=1}^{n}\partial _{j},L_{i+1}\right] &=&\left[
\sum\nolimits_{j=1}^{n}\partial _{j},\left[ \theta ,L_{i}\right] \right] =%
\left[ \left[ \sum\nolimits_{j=1}^{n}\partial _{j},\theta \right] ,L_{i}%
\right] +\left[ \left[ L_{i},\sum\nolimits_{j=1}^{n}\partial _{j}\right]
,\theta \right] \\
&=&\left[ \left( m^{-1}+n-1\right) Id,L_{i}^{m}\right] +\left[ 0,\theta %
\right] =0.
\end{eqnarray*}

To prove that operators $L_{k}$ and $L_{l},$ $k\neq l,$ commute one can use
the arguments of the paper \cite{O1998}. Consider an involution $\delta $ on
the space of all differential operators on $R^{n}$ corresponding to the
change $x\rightarrow -x$ and the standard anti-involution $\ast $: operator $%
L^{\ast }$ is a formal adjoint to $L.$ We have $\left[ L_{1}^{\delta
},L_{2}^{\delta }\right] =\left[ L_{1},L_{2}\right] ^{\delta }$ and $\left[
L_{1}^{\ast },L_{2}^{\ast }\right] =-\left[ L_{1},L_{2}\right] ^{\ast }.$
Operators $L_{k}$ have the following properties with respect to these
involutions: $L_{k}^{\ast }=L_{k}^{\delta }=\left( -1\right) ^{k}L_{k}.$
Now, consider the commutator $C=\left[ L_{k},L_{l}\right] $. By the Jacobi
identity $\left[ C,H\right] =0,$ therefore we can use Berezin's lemma~\cite%
{B1971} which states that in such a case the highest symbol of $C$ is
polynomial in $x.$ In this case it is also periodic, hence the highest
symbol must be constant. Now, we also have%
\begin{equation}
C^{\delta }=\left[ L_{k},L_{l}\right] ^{\delta }=\left[ L_{k}^{\delta
},L_{l}^{\delta }\right] =\left[ L_{k}^{\ast },L_{l}^{\ast }\right] =-\left[
L_{k},L_{l}\right] ^{\ast }=-C^{\ast }.  \label{C}
\end{equation}%
Since the highest symbol of $C$ is constant the highest symbols of $C^{\ast
} $ and $C^{\delta }$ are the same but it follows from (\ref{C}) they they
must be different by a sign. This means that the highest symbol of $C$ is
zero and hence $C$ is zero.

\textbf{Theorem 3.} \textit{Deformed quantum CM problem (\ref{operator}) is
integrable for all $n$ and $m$ and algebraically integrable for integer $m$.}

\textbf{Proof.} The complete family of the commuting quantum integrals for
arbitrary $m$ is given by the previous theorems. The algebraic integrability
in the case when $m$ is integer follows from the general result due to
Chalykh, Etingof and Oblomkov (see Theorem 3.8 in \cite{CEO2003}).

\section{Trigonometric and rational degenerations}

Trigonometric degenerations of the Weierstrass $\wp $-function corresponds
to the case when one of the half periods $\omega _{1}$ or $\omega _{2}$ is
infinite, which happens when two of the roots of the polynomial (\ref%
{differ_equation_P}) collide. For example, the case of $e_{1}=e_{2}=a$\ and $%
e_{3}=-2a$ corresponds to $\omega =\infty ,$ $\tilde{\omega}=i\frac{\pi }{%
\sqrt{12a}}$ and $\wp \left( z\right) =a+\frac{3a}{\sinh ^{2}\sqrt{3a}z}.$
Choosing $a=\frac{1}{3}$ we have%
\begin{equation*}
\begin{array}{c}
\wp \left( z\right) =\frac{1}{3}+\frac{1}{\sinh ^{2}z}=z^{-2}-\sum_{k=1}^{%
\infty }\frac{2^{2k+2}}{\left( 2k+2\right) }\frac{B_{2k+2}}{\left( 2k\right)
!}z^{2k},\text{\quad }\varsigma \left( z\right) =-\frac{1}{3}z+\coth z,\text{%
\quad and\quad }\gamma _{2k}=-\frac{2^{2k+2}}{\left( 2k+2\right) }\frac{%
B_{2k+2}}{\left( 2k\right) !},%
\end{array}%
\end{equation*}%
where $B_{2k+2}$ are the classical Bernoulli numbers defined by the expansion%
\begin{equation*}
\begin{array}{c}
\frac{z}{e^{z}-1}=1-\frac{1}{2}z+\sum_{k=1}^{\infty }\frac{B_{2k}}{\left(
2k\right) !}z^{2k}.%
\end{array}%
\end{equation*}%
In this case the Hamiltonian $H$\ takes the form%
\begin{equation*}
\begin{array}{c}
H=-\left( m\partial _{1}^{2}+\partial _{2}^{2}+\ldots +\partial
_{n-1}^{2}+\partial _{n}^{2}\right) +\frac{\left( m+1\right) \left(
n-1\right) }{3}\left( 1-m+\frac{mn}{2}\right) +\sum_{k=2}^{n}\frac{2\left(
m+1\right) }{\sinh ^{2}\left( x_{1}-x_{k}\right) }+\sum_{2\leq j<k\leq n}%
\frac{2m\left( m+1\right) }{\sinh ^{2}\left( x_{j}-x_{k}\right) }.%
\end{array}%
\end{equation*}%
Therefore, the formulae for the integrals in this case can be obtained using
the following recursive formulae for constants $p_{2i,k}:$ 
\begin{equation*}
\begin{array}{c}
p_{2i,k}=\frac{\left( 1-m\left( k-2i-1\right) \right) }{k-2i}%
p_{2i,k-1}+\left( 1+m\right) \sum\limits_{\substack{ j=1  \\ j\neq 2}}%
^{i-1}C_{k-2j+1}^{k-2i}\frac{2^{2i-2j+2}}{\left( 2i-2j+2\right) }%
B_{2i-2j+2}p_{2j-2,k-1}.%
\end{array}%
\end{equation*}%
The integrability of this problem was shown in \cite{CFV1998}. In \cite%
{SV2004} a recurrent formula was found for the quantum integrals with the
highest symbols given by the deformed Newton sums. Our formulae correspond
to the deformed elementary symmetric polynomials and seem to be new even in
that degenerate case.

The rational degeneration corresponds to both periods be equal to infinity.
In this case $\wp \left( z\right) =z^{-2}$ and all $\gamma _{2k}=0.$
Therefore, in this case only constants $p_{0,k},$ $k=1,2,\ldots ,$ are non
zero.

\section{Proof of Theorem 1.}

We prove Theorem 1 by induction in $n$. For small $n$ we showed this in the
section 3. Now assume that the statement of the theorem is true for all $k<n$
and show that it is true for $k=n.$

Under this assumption, let us first show that commutator $\left[ I,H\right] $
can be reduced to an expression on the additional terms $X,$ $X_{\hat{\jmath}%
}$ and $X_{\hat{k}\hat{l}}.$

\textbf{Lemma 1.}%
\begin{equation}
\left[ I,H\right] =\left[ X,H\right] +\sum\nolimits_{j=1}^{n}\left[ X_{\hat{%
\jmath}}\partial _{j},2\sum\nolimits_{\substack{ l=1  \\ l\neq j}}^{n}u_{jl}%
\right] +\sum\nolimits_{1\leq k<l\leq n}\left[ X_{\hat{k}\hat{l}}\partial
_{k}\partial _{l},2u_{kl}\right] .  \label{tails}
\end{equation}

\textbf{Proof. }Using (\ref{I}) we obtain%
\begin{equation*}
\left[ I,H\right] =\sum_{t=1}^{n-2}\left( -1\right) ^{t+1}\sum_{\sigma \in 
\mathfrak{S}\left( t\right) }\left[ I_{\hat{\sigma}}\partial _{\sigma },H%
\right] +\left( -1\right) ^{n}\left( n-1\right) \left[ \partial _{1}\ldots
\partial _{n},H\right] +\left[ X,H\right] ,
\end{equation*}%
which is equal to%
\begin{eqnarray*}
&&\sum_{j=1}^{n}\left[ I_{\hat{\jmath}}\partial _{j},H_{\hat{\jmath}}-\Delta
_{j}+2\sum\limits_{\substack{ k=1 \\ k\neq j}}^{n}u_{jk}\right]
+\sum_{t=2}^{n-2}\left( -1\right) ^{t+1}\sum_{\sigma \in \mathfrak{S}\left(
t\right) }\left[ I_{\hat{\sigma}}\partial _{\sigma },H_{\hat{\sigma}}-\Delta
_{\sigma }+2\sum\limits_{j\in \sigma }\sum\limits_{\substack{ k=1 \\ k\neq j
}}^{n}u_{jk}-2\sum\limits_{\substack{ j<k \\ j,k\in \sigma }}u_{jk}\right] 
\\
&&+\left( -1\right) ^{n}\left( n-1\right) \left[ \partial _{1}\ldots
\partial _{n},2\sum\limits_{1\leq j<k\leq n}u_{jk}\right] +\left[ X,H\right]
.
\end{eqnarray*}%
Since $I_{\hat{\sigma}}$ commutes with $H_{\hat{\sigma}}$ (by the induction
assumption) and with $-\Delta _{\sigma }$ (since it does not depend on $%
x_{i},$ $i\in \sigma )$ we have%
\begin{eqnarray*}
\left[ I,H\right]  &=&\sum_{j=1}^{n}\left[ I_{\hat{\jmath}}\partial
_{j},2\sum\limits_{\substack{ k=1 \\ k\neq j}}^{n}u_{jk}\right]
+\sum_{t=2}^{n-2}\left( -1\right) ^{t+1}\sum_{\sigma \in \mathfrak{S}\left(
t\right) }\left[ I_{\hat{\sigma}}\partial _{\sigma },2\sum\limits_{j\in
\sigma }\sum\limits_{\substack{ k=1 \\ k\neq j}}^{n}u_{jk}-2\sum\limits
_{\substack{ j<k \\ j,k\in \sigma }}u_{jk}\right]  \\
&&+\left( -1\right) ^{n}\left( n-1\right) \left[ \partial _{1}\ldots
\partial _{n},2\sum\limits_{1\leq j<k\leq n}u_{jk}\right] +\left[ X,H\right]
.
\end{eqnarray*}%
Now we use (\ref{I}) again and obtain%
\begin{eqnarray*}
\left[ I,H\right]  &=&\sum_{j=1}^{n}\left[ \sum_{t=1}^{n-3}\left( -1\right)
^{t+1}\sum_{\sigma \in \mathfrak{S}\left( \hat{\jmath};t\right) }I_{\hat{%
\jmath}\hat{\sigma}}\partial _{\sigma }\partial _{j},2\sum\limits_{\substack{
k=1 \\ k\neq j}}^{n}u_{jk}\right] +\sum_{j=1}^{n}\left[ \left( -1\right)
^{n-1}\left( n-2\right) \partial _{1}\ldots \partial _{n}+X_{\hat{\jmath}%
}\partial _{j},2\sum\limits_{\substack{ k=1 \\ k\neq j}}^{n}u_{jk}\right]  \\
&&+\sum_{t=2}^{n-2}\left( -1\right) ^{t+1}\sum_{\sigma \in \mathfrak{S}%
\left( t\right) }\left[ I_{\hat{\sigma}}\partial _{\sigma
},2\sum\limits_{j\in \sigma }\sum\limits_{\substack{ k=1 \\ k\neq j}}%
^{n}u_{jk}-2\sum\limits_{\substack{ j<k \\ j,k\in \sigma }}u_{jk}\right]  \\
&&+\left( -1\right) ^{n}\left( n-1\right) \left[ \partial _{1}\ldots
\partial _{n},2\sum\limits_{1\leq j<k\leq n}u_{jk}\right] +\left[ X,H\right]
.
\end{eqnarray*}%
If we cancel the repeated terms we obtain%
\begin{eqnarray*}
\left[ I,H\right]  &=&\sum_{j=1}^{n}\left[ \left( -1\right) ^{n-1}\left(
n-2\right) \partial _{1}\ldots \partial _{n}+X_{\hat{\jmath}}\partial
_{j},2\sum\limits_{\substack{ k=1 \\ k\neq j}}^{n}u_{jk}\right]
+\sum_{t=2}^{n-2}\left( -1\right) ^{t}\sum_{\sigma \in \mathfrak{S}\left(
t\right) }\left[ I_{\hat{\sigma}}\partial _{\sigma },2\sum\limits_{\substack{
j<k \\ j,k\in \sigma }}u_{jk}\right]  \\
&&+\left( -1\right) ^{n}\left( n-1\right) \left[ \partial _{1}\ldots
\partial _{n},2\sum\limits_{1\leq j<k\leq n}u_{jk}\right] +\left[ X,H\right]
.
\end{eqnarray*}%
We use (\ref{I}) again%
\begin{eqnarray*}
\left[ I,H\right]  &=&\sum_{j=1}^{n}\left[ \left( -1\right) ^{n-1}\left(
n-2\right) \partial _{1}\ldots \partial _{n}+X_{\hat{\jmath}}\partial
_{j},2\sum\limits_{\substack{ k=1 \\ k\neq j}}^{n}u_{jk}\right]
+\sum\limits_{1\leq j<k\leq n}\left[ \sum_{t=1}^{n-4}\left( -1\right)
^{t+1}\sum_{\sigma \in \mathfrak{S}\left( \hat{\jmath}\hat{k};t\right) }I_{%
\hat{\sigma}\hat{\jmath}\hat{k}}\partial _{\sigma }\partial _{j}\partial
_{k},2u_{jk}\right]  \\
&&+\sum\limits_{1\leq j<k\leq n}\left[ \left( -1\right) ^{n-2}\left(
n-3\right) \partial _{1}\ldots \partial _{n}+X_{\hat{\jmath}\hat{k}}\partial
_{j}\partial _{k},2u_{jk}\right] +\sum_{t=3}^{n-2}\left( -1\right)
^{t}\sum_{\sigma \in \mathfrak{S}\left( t\right) }\left[ I_{\hat{\sigma}%
}\partial _{\sigma },2\sum\limits_{\substack{ j<k \\ j,k\in \sigma }}u_{jk}%
\right]  \\
&&+\left( -1\right) ^{n}\left( n-1\right) \left[ \partial _{1}\ldots
\partial _{n},2\sum\limits_{1\leq j<k\leq n}u_{jk}\right] +\left[ X,H\right]
.
\end{eqnarray*}%
Finally, canceling the repeated terms, we get%
\begin{eqnarray*}
\left[ I,H\right]  &=&\left( -1\right) ^{n}2\left[ \partial _{1}\ldots
\partial _{n},-\left( n-2\right) \sum_{j=1}^{n}\sum\limits_{\substack{ k=1
\\ k\neq j}}^{n}u_{jk}+\left( n-3\right) \sum\limits_{1\leq j<k\leq
n}u_{jk}+\left( n-1\right) \sum\limits_{1\leq j<k\leq n}u_{jk}\right]  \\
&&+\left[ X,H\right] +\sum_{j=2}^{n}\left[ X_{\hat{\jmath}}\partial
_{j},2\sum\limits_{\substack{ k=1 \\ k\neq j}}^{n}u_{jk}\right]
+\sum\limits_{1\leq j<k\leq n}\left[ X_{\hat{\jmath}\hat{k}}\partial
_{j}\partial _{k},2u_{jk}\right] ,
\end{eqnarray*}%
which simplifies to%
\begin{equation*}
\left[ I,H\right] =\left[ X,H\right] +\sum_{j=1}^{n}\left[ X_{\hat{\jmath}%
}\partial _{j},2\sum\limits_{\substack{ k=1 \\ k\neq j}}^{n}u_{jk}\right]
+\sum\limits_{1\leq j<k\leq n}\left[ X_{\hat{\jmath}\hat{k}}\partial
_{j}\partial _{k},2u_{jk}\right] .
\end{equation*}%
Lemma 1 is proven.

\textbf{Lemma 2. }In the non deformed case%
\begin{equation}
\left[ I_{\hat{1}},H_{\hat{1}}\right] =\left[ X_{\hat{1}},H_{\hat{1}}\right]
+\sum\limits_{j=2}^{n}\left[ X_{\hat{1}\hat{\jmath}}\partial
_{j},2\sum\limits_{\substack{ l=2  \\ l\neq j}}^{n}u_{jl}\right]
+\sum\limits_{2\leq k<l\leq n}\left[ X_{\hat{1}\hat{k}\hat{l}}\partial
_{k}\partial _{l},2u_{kl}\right] =0.  \label{X_no_difformation}
\end{equation}

\textbf{Proof.} To prove that 
\begin{equation*}
\left[ I_{\hat{1}},H_{\hat{1}}\right] =\left[ X_{\hat{1}},H_{\hat{1}}\right]
+\sum\limits_{j=2}^{n}\left[ X_{\hat{1}\hat{\jmath}}\partial
_{j},2\sum\limits_{\substack{ l=2  \\ l\neq j}}^{n}u_{jl}\right]
+\sum\limits_{2\leq k<l\leq n}\left[ X_{\hat{1}\hat{k}\hat{l}}\partial
_{k}\partial _{l},2u_{kl}\right]
\end{equation*}%
one must repeat the arguments of Lemma 1. We will need the addition theorem
for the Weierstrass elliptic function~\cite{WW1927}%
\begin{equation*}
T_{ijk}\equiv \det \left( 
\begin{array}{ccc}
\wp _{ij} & \wp _{jk} & \wp _{ki} \\ 
\wp _{ij}^{\prime } & \wp _{jk}^{\prime } & \wp _{ki}^{\prime } \\ 
1 & 1 & 1%
\end{array}%
\right) \equiv 0.
\end{equation*}%
Two cases must be considered to prove the lemma: $n$ is odd and $n$ is even.
If $n$ is even (\ref{X_no_difformation}) is reduced to $2\sum%
\limits_{j=2}^{n}\sum\limits_{\substack{ l=2  \\ l\neq j}}^{n}\left[ X_{\hat{%
1}\hat{\jmath}}\partial _{j},u_{jl}\right] =0,$ as $X_{\hat{1}}=0$ and $X_{%
\hat{1}\hat{k}\hat{l}}=0,$ $2\leq k<l\leq n.$ In this case we have%
\begin{eqnarray*}
\sum\limits_{j=2}^{n}\sum\limits_{\substack{ l=2  \\ l\neq j}}^{n}\left[ X_{%
\hat{1}\hat{\jmath}}\partial _{j},u_{jl}\right] &=&\sum\limits_{j=2}^{n}\sum%
\limits_{\substack{ l=2  \\ l\neq j}}^{n}X_{\hat{1}\hat{\jmath}}\left[
\partial _{j},u_{jl}\right] =m\left( m+1\right)
\sum\limits_{j=2}^{n}\sum\limits_{\substack{ l=2  \\ l\neq j}}^{n}X_{\hat{1}%
\hat{\jmath}}\wp _{jl}^{\prime } \\
&=&m\left( m+1\right) \sum\limits_{2\leq j<l\leq n}\wp _{jl}^{\prime }\left(
X_{\hat{1}\hat{\jmath}}-X_{\hat{1}\hat{l}}\right) =m\left( m+1\right)
\sum\limits_{2\leq i<j<k\leq n}T_{ijk}X_{\hat{1}\hat{\imath}\hat{\jmath}\hat{%
k}}=0.
\end{eqnarray*}%
If $n$ is odd (\ref{X_no_difformation}) becomes $\left[ X_{\hat{1}},-\Delta
_{\hat{1}}\right] +\sum\limits_{2\leq k<l\leq n}\left[ X_{\hat{1}\hat{k}\hat{%
l}}\partial _{k}\partial _{l},2u_{kl}\right] =0,$ since in this case $X_{%
\hat{1}\hat{\jmath}}=0,$ $j=2,\ldots ,n.$ We have%
\begin{eqnarray*}
\left[ X_{\hat{1}},-\Delta _{\hat{1}}\right] +\sum\limits_{2\leq k<l\leq n}%
\left[ X_{\hat{1}\hat{k}\hat{l}}\partial _{k}\partial _{l},2u_{kl}\right]
&=&\sum\limits_{k=2}^{n}\frac{\partial ^{2}X_{\hat{1}}}{\partial x_{k}^{2}}%
+2\sum\limits_{k=2}^{n}\frac{\partial X_{\hat{1}}}{\partial x_{k}}\partial
_{k}+2\sum\limits_{2\leq k<l\leq n}X_{\hat{1}\hat{k}\hat{l}}\left(
u_{kl}^{\prime }\partial _{l}-u_{kl}^{\prime }\partial _{k}-u_{kl}^{\prime
\prime }\right) \\
&=&\sum\limits_{k=2}^{n}\frac{\partial ^{2}X_{\hat{1}}}{\partial x_{k}^{2}}%
-2\sum\limits_{2\leq k<l\leq n}X_{\hat{1}\hat{k}\hat{l}}u_{kl}^{\prime
\prime }+2\sum\limits_{k=2}^{n}\left( \frac{\partial X_{\hat{1}}}{\partial
x_{k}}-\sum\limits_{\substack{ l=2  \\ l\neq k}}^{n}X_{\hat{1}\hat{k}\hat{l}%
}u_{kl}^{\prime }\right) \partial _{k}=0.
\end{eqnarray*}%
Lemma 2 is proven.

\textbf{Lemma 3. }The operators $\mathfrak{D}^{n}$ satisfy the relation%
\begin{equation}
\begin{array}{l}
\left[ \mathfrak{D}^{n},\wp \left( x_{1}-x_{i}\right) \right] +\left(
1+m\right) \left[ \left[ \varsigma \left( x_{1}-x_{i}\right) ,\mathfrak{D}%
^{n-1}\right] ,\wp \left( x_{1}-x_{i}\right) \right] \\ 
\qquad +m\left[ \mathfrak{D}^{n-1},\wp \left( x_{1}-x_{i}\right) \right]
\partial -\frac{\left( 1+m\right) }{2}\wp ^{\prime }\left(
x_{1}-x_{i}\right) \mathfrak{D}^{n-1}-\frac{\left( 1-m\right) }{2}\mathfrak{D%
}^{n-1}\wp ^{\prime }\left( x_{1}-x_{i}\right) =0%
\end{array}
\label{D}
\end{equation}%
for any $i=2,\ldots ,n,$ where $\partial =\frac{\partial }{\partial x_{1}}.$

\textbf{Proof.}

Let us denote the left hand side of the relation (\ref{D}) by $\mathcal{Y}%
_{n}.$

It can be shown by a simple direct calculation that (\ref{D}) is true if $%
n=1,2,3,4.$ The equations in these cases are%
\begin{equation*}
\begin{array}{l}
n=1:\mathcal{Y}_{1}=\left[ \partial ,\wp \right] +m\left[ 1,\wp \right]
\partial -\frac{\left( 1+m\right) }{2}\wp ^{\prime }-\frac{\left( 1-m\right) 
}{2}\wp ^{\prime }=0, \\ 
n=2:\mathcal{Y}_{2}=\left[ \frac{1-m}{2}\partial ^{2},\wp \right] +\left(
1+m\right) \left[ \left[ \varsigma ,\partial \right] ,\wp \right] +m\left[
\partial ,\wp \right] \partial -\frac{\left( 1+m\right) }{2}\wp ^{\prime
}\partial -\frac{\left( 1-m\right) }{2}\partial \wp ^{\prime }=0, \\ 
n=3:\mathcal{Y}_{3}=\left[ \frac{1-2m}{3}\partial ^{3},\wp \right] +\left(
1+m\right) \left[ \left[ \varsigma ,\partial ^{2}\right] ,\wp \right] +m%
\left[ \partial ^{2},\wp \right] \partial -\frac{\left( 1+m\right) }{2}\wp
^{\prime }\partial ^{2}-\frac{\left( 1-m\right) }{2}\partial ^{2}\wp
^{\prime }=0, \\ 
n=4:\mathcal{Y}_{4}=\left[ \frac{1-3m}{4}\partial ^{4},\wp \right] +\left(
1+m\right) \left[ \left[ \varsigma ,\partial ^{3}\right] ,\wp \right] +m%
\left[ \partial ^{3},\wp \right] \partial -\frac{\left( 1+m\right) }{2}\wp
^{\prime }\partial ^{3}-\frac{\left( 1-m\right) }{2}\partial ^{3}\wp
^{\prime }=0.%
\end{array}%
\end{equation*}%
For $n\geq 5$ we use (\ref{D_formulae}) to obtain%
\begin{eqnarray*}
\mathcal{Y}_{n} &=&\left[ p_{0,n}\partial ^{n},\wp \right] +\left(
1+m\right) \left[ \left[ \varsigma ,p_{0,n-1}\partial ^{n-1}\right] ,\wp %
\right] +m\left[ p_{0,n-1}\partial ^{n-1},\wp \right] \partial -\frac{\left(
1+m\right) }{2}\wp ^{\prime }p_{0,n-1}\partial ^{n-1} \\
&&%
\begin{array}{c}
-\frac{\left( 1-m\right) }{2}p_{0,n-1}\partial ^{n-1}\wp ^{\prime }+\left[
\sum\limits_{i=2}^{\left[ \frac{n}{2}\right] }p_{2i,n}\partial ^{n-2i},\wp %
\right] +\left( 1+m\right) \left[ \left[ \varsigma ,\sum\limits_{i=2}^{\left[
\frac{n-1}{2}\right] }p_{2i,n-1}\partial ^{n-1-2i}\right] ,\wp \right]%
\end{array}
\\
&&%
\begin{array}{c}
+m\left[ \sum\limits_{i=2}^{\left[ \frac{n-1}{2}\right] }p_{2i,n-1}\partial
^{n-1-2i},\wp \right] \partial -\frac{\left( 1+m\right) }{2}\wp ^{\prime
}\sum\limits_{i=2}^{\left[ \frac{n-1}{2}\right] }p_{2i,n-1}\partial
^{n-1-2i}-\frac{\left( 1-m\right) }{2}\sum\limits_{i=2}^{\left[ \frac{n-1}{2}%
\right] }p_{2i,n-1}\partial ^{n-1-2i}\wp ^{\prime }.%
\end{array}%
\end{eqnarray*}%
We can express $p_{0,n}$ through $p_{0,n-1}$ and $p_{2i,n}$ through $%
p_{2l,n-1},$ $l=0,\ldots 2i,$ as described by (\ref{p_2i,k}) and obtain%
\begin{eqnarray*}
\mathcal{Y}_{n} &=&%
\begin{array}{c}
\left[ \frac{1-m\left( n-1\right) }{n}p_{0,n-1}\partial ^{n},\wp \right]
+\left( 1+m\right) \left[ \left[ \varsigma ,p_{0,n-1}\partial ^{n-1}\right]
,\wp \right] +m\left[ p_{0,n-1}\partial ^{n-1},\wp \right] \partial%
\end{array}
\\
&&%
\begin{array}{c}
-\frac{\left( 1+m\right) }{2}\wp ^{\prime }p_{0,n-1}\partial ^{n-1}-\frac{%
\left( 1-m\right) }{2}p_{0,n-1}\partial ^{n-1}\wp ^{\prime }-\left(
1+m\right) \left[ \sum\limits_{i=2}^{\left[ \frac{n-1}{2}\right] }\left(
2i-2\right) !C_{n-1}^{n-2i}\gamma _{2i-2}p_{0,n-1}\partial ^{n-2i},\wp %
\right]%
\end{array}
\\
&&%
\begin{array}{c}
+\left[ \sum\limits_{i=2}^{\left[ \frac{n-1}{2}\right] }\frac{\left(
1-m\left( n-2i-1\right) \right) }{n-2i}p_{2i,n-1}\partial ^{n-2i},\wp \right]
+\left( 1+m\right) \left[ \left[ \varsigma ,\sum\limits_{i=2}^{\left[ \frac{%
n-1}{2}\right] }p_{2i,n-1}\partial ^{n-1-2i}\right] ,\wp \right]%
\end{array}
\\
&&%
\begin{array}{c}
+m\left[ \sum\limits_{i=2}^{\left[ \frac{n-1}{2}\right] }p_{2i,n-1}\partial
^{n-1-2i},\wp \right] \partial -\frac{\left( 1+m\right) }{2}\wp ^{\prime
}\sum\limits_{i=2}^{\left[ \frac{n-1}{2}\right] }p_{2i,n-1}\partial
^{n-1-2i}-\frac{\left( 1-m\right) }{2}\sum\limits_{i=2}^{\left[ \frac{n-1}{2}%
\right] }p_{2i,n-1}\partial ^{n-1-2i}\wp ^{\prime }%
\end{array}
\\
&&%
\begin{array}{c}
-\left( 1+m\right) \sum\limits_{i=2}^{\left[ \frac{n-1}{2}\right] -2}\left[
\sum\limits_{j=i+2}^{\left[ \frac{n-1}{2}\right] }\left( 2j-2i-2\right)
!C_{n-2i-1}^{n-2j}\gamma _{2j-2i-2}p_{2i,n-1}\partial ^{n-2j},\wp \right] .%
\end{array}%
\end{eqnarray*}%
$\allowbreak $Denote%
\begin{eqnarray*}
\mathcal{W}_{k} &=&%
\begin{array}{c}
\left[ \frac{1-m\left( k-1\right) }{k}\partial ^{k},\wp \right] +\left(
1+m\right) \left[ \left[ \varsigma ,\partial ^{k-1}\right] ,\wp \right] +m%
\left[ \partial ^{k-1},\wp \right] \partial%
\end{array}
\\
&&%
\begin{array}{c}
-\frac{\left( 1+m\right) }{2}\wp ^{\prime }\partial ^{k-1}-\frac{\left(
1-m\right) }{2}\partial ^{k-1}\wp ^{\prime }-\left( 1+m\right)
\sum\limits_{i=2}^{\left[ \frac{k-1}{2}\right] }\left[ \left( 2i-2\right)
!C_{k-1}^{k-2i}\gamma _{2i-2}\partial ^{k-2i},\wp \right] ,%
\end{array}%
\end{eqnarray*}%
then%
\begin{gather*}
\mathcal{Y}_{5}=p_{0,4}\mathcal{W}_{5}+p_{4,4}\mathcal{Y}_{1},\quad \mathcal{%
Y}_{6}=p_{0,5}\mathcal{W}_{6}+p_{4,5}\mathcal{Y}_{2}, \\
\mathcal{Y}_{7}=p_{0,6}\mathcal{W}_{7}+p_{4,6}\mathcal{Y}_{3}+p_{6,6}%
\mathcal{Y}_{1},\quad \mathcal{Y}_{8}=p_{0,7}\mathcal{W}_{8}+p_{4,7}\mathcal{%
Y}_{4}+p_{6,7}\mathcal{Y}_{2}
\end{gather*}%
and for $k\geq 5$%
\begin{eqnarray*}
\mathcal{Y}_{2k-1} &=&p_{0,2k-2}\mathcal{W}_{2k-1}+\sum%
\limits_{i=2}^{k-3}p_{2i,2k-2}\mathcal{W}_{2k-1-2i}+p_{2k-4,2k-2}\mathcal{Y}%
_{3}+p_{2k-2,2k-2}\mathcal{Y}_{1}, \\
\mathcal{Y}_{2k} &=&p_{0,2k-1}\mathcal{W}_{2k}+\sum%
\limits_{i=2}^{k-3}p_{2i,2k-1}\mathcal{W}_{2k-2i}+p_{2k-4,2k-1}\mathcal{Y}%
_{4}+p_{2k-2,2k-1}\mathcal{Y}_{2}.
\end{eqnarray*}%
Below we show that $\mathcal{W}_{n}=0$ for $n\geq 5.$ Firstly,%
\begin{eqnarray*}
\begin{array}{c}
\frac{10}{1+m}\mathcal{W}_{5}%
\end{array}
&=&%
\begin{array}{c}
2\left[ \partial ^{4},\wp \right] \partial -3\partial ^{4}\wp ^{\prime }+10%
\left[ \left[ \varsigma ,\partial ^{4}\right] ,\wp \right] -5\wp ^{\prime
}\partial ^{4}-80\left[ \gamma _{2}\partial ^{1},\wp \right] +2\partial %
\left[ \partial ^{3},\wp \right] \partial%
\end{array}
\\
&&%
\begin{array}{c}
-3\wp ^{\prime }\partial ^{4}-3\partial ^{4}\wp ^{\prime }+10\partial \left[ %
\left[ \varsigma ,\partial ^{3}\right] ,\wp \right] +10\left[ \partial ,\wp %
\right] \left[ \varsigma ,\partial ^{3}\right] +10\left[ \varsigma ,\partial %
\right] \left[ \partial ^{3},\wp \right] -80\left[ \gamma _{2}\partial
^{1},\wp \right] .%
\end{array}%
\end{eqnarray*}%
We calculate that $\partial \left[ \left[ \varsigma ,\partial ^{3}\right]
,\wp \right] =-\frac{1}{4}\partial \left[ \partial ^{3},\wp \right] \partial
+\frac{1}{2}\partial \wp ^{\prime }\partial ^{3}+\frac{1}{4}\partial ^{4}\wp
^{\prime },$ therefore%
\begin{eqnarray*}
\begin{array}{c}
\frac{10}{1+m}\mathcal{W}_{5}%
\end{array}
&=&%
\begin{array}{c}
-\frac{1}{2}\partial \left[ \partial ^{3},\wp \right] \partial -3\wp
^{\prime }\partial ^{4}-\frac{1}{2}\partial ^{4}\wp ^{\prime }+5\partial \wp
^{\prime }\partial ^{3}+10\wp ^{\prime }\left[ \varsigma ,\partial ^{3}%
\right] +10\wp \left[ \partial ^{3},\wp \right] -80\gamma _{2}\wp ^{\prime }%
\end{array}
\\
&&%
\begin{array}{c}
-5\wp ^{\prime \prime \prime }\partial ^{2}-\frac{5}{2}\wp ^{\left( 4\right)
}\partial -\frac{1}{2}\wp ^{\left( 5\right) }-80\gamma _{2}\wp ^{\prime }%
\end{array}
\\
&&%
\begin{array}{c}
+10\wp ^{\prime }\left( 3\wp \partial ^{2}+3\wp ^{\prime }\partial +\wp
^{\prime \prime }\right) +10\wp \left( 3\wp ^{\prime }\partial ^{2}+3\wp
^{\prime \prime }\partial +\wp ^{\prime \prime \prime }\right) =0.%
\end{array}%
\end{eqnarray*}%
In this calculation the identities which are the derivatives of the
differential equation of the Weierstrass elliptic function (\ref%
{differ_equation_P}) have been used. At the next step we use the induction
assumption that for any $s<k$ $\mathcal{W}_{s}=0.$\ Then $\mathcal{W}_{k}$
can be simplified into%
\begin{eqnarray*}
\mathcal{W}_{k} &=&%
\begin{array}{c}
\frac{1+m}{k}\left[ \partial ^{k-1},\wp \right] \partial -\frac{\left(
1+m\right) }{2}\frac{k-2}{k}\partial ^{k-1}\wp ^{\prime }+\left( 1+m\right) %
\left[ \left[ \varsigma ,\partial ^{k-1}\right] ,\wp \right]%
\end{array}
\\
&&%
\begin{array}{c}
-\frac{\left( 1+m\right) }{2}\wp ^{\prime }\partial ^{k-1}-\left( 1+m\right)
\sum\limits_{j=2}^{\left[ \frac{k-1}{2}\right] }\left[ \left( 2j-2\right)
!C_{k-1}^{k-2j}\gamma _{2j-2}\partial ^{k-2j},\wp \right] .%
\end{array}%
\end{eqnarray*}%
We consider%
\begin{eqnarray*}
\begin{array}{c}
\frac{1}{1+m}\mathcal{W}_{k}%
\end{array}
&=&%
\begin{array}{c}
\frac{1}{k}\left[ \partial ^{k-1},\wp \right] \partial -\frac{k-2}{2k}%
\partial ^{k-1}\wp ^{\prime }+\left[ \left[ \varsigma ,\partial ^{k-1}\right]
,\wp \right] -\frac{1}{2}\wp ^{\prime }\partial ^{k-1}%
\end{array}
\\
&&%
\begin{array}{c}
-\sum\limits_{j=2}^{\left[ \frac{k-1}{2}\right] }\left[ \left( 2j-2\right)
!C_{k-1}^{k-2j}\gamma _{2j-2}\partial ^{k-2j},\wp \right] .%
\end{array}%
\end{eqnarray*}%
If $k=2l$%
\begin{eqnarray*}
\begin{array}{c}
\frac{1}{1+m}\mathcal{W}_{2l}%
\end{array}
&=&%
\begin{array}{c}
\frac{1}{2l}\left[ \partial ^{2l-1},\wp \right] \partial -\frac{l-1}{2l}%
\partial ^{2l-1}\wp ^{\prime }+\left[ \left[ \varsigma ,\partial ^{2l-1}%
\right] ,\wp \right] -\frac{1}{2}\wp ^{\prime }\partial ^{2l-1}%
\end{array}
\\
&&%
\begin{array}{c}
-\sum\limits_{j=2}^{l-1}\left[ \left( 2j-2\right) !C_{2l-1}^{2l-2j}\gamma
_{2j-2}\partial ^{2l-2j},\wp \right]%
\end{array}%
\end{eqnarray*}%
and if $k=2l+1$%
\begin{eqnarray*}
\begin{array}{c}
\frac{1}{1+m}\mathcal{W}_{2l+1}%
\end{array}
&=&%
\begin{array}{c}
\frac{1}{\left( 2l+1\right) }\left[ \partial ^{2l},\wp \right] \partial -%
\frac{2l-1}{2\left( 2l+1\right) }\partial ^{2l}\wp ^{\prime }+\left[ \left[
\varsigma ,\partial ^{2l}\right] ,\wp \right] -\frac{1}{2}\wp ^{\prime
}\partial ^{2l}%
\end{array}
\\
&&%
\begin{array}{c}
-\sum\limits_{j=2}^{l}\left[ \left( 2j-2\right) !C_{2l}^{2l+1-2j}\gamma
_{2j-2}\partial ^{2l+1-2j},\wp \right] .%
\end{array}%
\end{eqnarray*}%
We show below that $\mathcal{W}_{2l}=0.$ The case of $\mathcal{W}_{2l+1}$
can be dealt with in the same way.%
\begin{eqnarray*}
\begin{array}{c}
\frac{1}{1+m}\mathcal{W}_{2l}%
\end{array}
&=&%
\begin{array}{c}
\frac{1}{2l}\left[ \partial ^{2l-1},\wp \right] \partial -\frac{1}{2l}\left(
l-1\right) \partial ^{2l-1}\wp ^{\prime }+\left[ \left[ \varsigma ,\partial
^{2l-1}\right] ,\wp \right]%
\end{array}
\\
&&%
\begin{array}{c}
-\frac{1}{2}\wp ^{\prime }\partial ^{2l-1}-\sum\limits_{j=2}^{l-1}\left[
\left( 2j-2\right) !C_{2l-1}^{2l-2j}\gamma _{2j-2}\partial ^{2l-2j},\wp %
\right]%
\end{array}
\\
&=&%
\begin{array}{c}
\frac{1}{2l}\left[ \partial \partial ^{2l-2},\wp \right] \partial -\frac{l-1%
}{2l}\partial \partial ^{2l-2}\wp ^{\prime }+\left[ \left[ \varsigma
,\partial \partial ^{2l-2}\right] ,\wp \right]%
\end{array}
\\
&&%
\begin{array}{c}
-\frac{1}{2}\wp ^{\prime }\partial \partial ^{2l-2}-\sum\limits_{j=2}^{l-1}%
\left[ \left( 2j-2\right) !C_{2l-1}^{2l-2j}\gamma _{2j-2}\partial \partial
^{2l-2j-1},\wp \right]%
\end{array}
\\
&=&%
\begin{array}{c}
\frac{1}{2l}\partial \left[ \partial ^{2l-2},\wp \right] \partial -\frac{l-1%
}{2l}\partial \partial ^{2l-2}\wp ^{\prime }-\frac{l-1}{2l}\wp ^{\prime
}\partial \partial ^{2l-2}+\partial \left[ \left[ \varsigma ,\partial ^{2l-2}%
\right] ,\wp \right]%
\end{array}
\\
&&%
\begin{array}{c}
+\left[ \partial ,\wp \right] \left[ \varsigma ,\partial ^{2l-2}\right] +%
\left[ \varsigma ,\partial \right] \left[ \partial ^{2l-2},\wp \right]%
\end{array}
\\
&&%
\begin{array}{c}
-\partial \sum\limits_{j=2}^{l-1}\left[ \left( 2j-2\right)
!C_{2l-1}^{2l-2j}\gamma _{2j-2}\partial ^{2l-2j-1},\wp \right]
-\sum\limits_{j=2}^{l-1}\left( 2j-2\right) !C_{2l-1}^{2l-2j}\gamma
_{2j-2}\wp ^{\prime }\partial ^{2l-2j-1}.%
\end{array}%
\end{eqnarray*}%
From the inductive assumption we have that $\mathcal{W}_{2l-1}=0,$ then%
\begin{eqnarray*}
\left[ \left[ \varsigma ,\partial ^{2l-2}\right] ,\wp \right] &=&%
\begin{array}{c}
-\frac{1}{2l-1}\left[ \partial ^{2l-2},\wp \right] \partial +\frac{2l-3}{%
2\left( 2l-1\right) }\partial ^{2l-2}\wp ^{\prime }+\frac{1}{2}\wp ^{\prime
}\partial ^{2l-2}%
\end{array}
\\
&&%
\begin{array}{c}
+\sum\limits_{j=2}^{l-1}\left[ \left( 2j-2\right) !C_{2l-2}^{2l-1-2j}\gamma
_{2j-2}\partial ^{2l-1-2j},\wp \right]%
\end{array}%
\end{eqnarray*}%
$\allowbreak $and therefore%
\begin{eqnarray*}
\begin{array}{c}
-\frac{1}{1+m}\mathcal{W}_{2l}%
\end{array}
&=&%
\begin{array}{c}
\frac{1}{2l\left( 2l-1\right) }\partial \left[ \partial ^{2l-2},\wp \right]
\partial +\frac{1}{2l\left( 2l-1\right) }\partial \partial ^{2l-2}\wp
^{\prime }-\frac{1}{2}\wp ^{\prime \prime }\partial ^{2l-2}-\frac{1}{2l}\wp
^{\prime }\partial \partial ^{2l-2}%
\end{array}
\\
&&%
\begin{array}{c}
+\partial \sum\limits_{j=2}^{l-1}\left[ \frac{\left( 2l-2\right) !}{\left(
2l-2j\right) !}\gamma _{2j-2}\partial ^{2l-1-2j},\wp \right] -\wp ^{\prime }%
\left[ \varsigma ,\partial ^{2l-2}\right] -\wp \left[ \partial ^{2l-2},\wp %
\right]%
\end{array}
\\
&&%
\begin{array}{c}
+\sum\limits_{j=2}^{l-1}\left( 2j-2\right) !C_{2l-1}^{2l-2j}\gamma
_{2j-2}\wp ^{\prime }\partial ^{2l-2j-1}%
\end{array}
\\
&&%
\begin{array}{c}
=\frac{1}{2l\left( 2l-1\right) }\partial \left[ \partial ^{2l-1},\wp \right]
-\frac{1}{2}\wp ^{\prime \prime }\partial ^{2l-2}-\frac{1}{2l}\wp ^{\prime
}\partial ^{2l-1}-\wp ^{\prime }\left[ \varsigma ,\partial ^{2l-2}\right]
-\wp \left[ \partial ^{2l-2},\wp \right]%
\end{array}
\\
&&%
\begin{array}{c}
+\partial \sum\limits_{j=2}^{l-1}\frac{\left( 2l-2\right) !}{\left(
2l-2j\right) !}\gamma _{2j-2}\left[ \partial ^{2l-1-2j},\wp \right]
+\sum\limits_{j=2}^{l-1}\left( 2j-2\right) !C_{2l-1}^{2l-2j}\gamma
_{2j-2}\wp ^{\prime }\partial ^{2l-2j-1}.%
\end{array}%
\end{eqnarray*}%
The commutators can be rewritten as%
\begin{eqnarray*}
\left[ \partial ^{2l-1},\wp \right] &=&\sum_{k=0}^{2l-2}C_{2l-1}^{2l-1-k}\wp
^{\left( 2l-1-k\right) }\partial ^{k},\qquad \left[ \partial ^{2l-2},\wp %
\right] =\sum_{k=0}^{2l-3}C_{2l-2}^{2l-2-k}\wp ^{\left( 2l-2-k\right)
}\partial ^{k}, \\
\left[ \varsigma ,\partial ^{2l-2}\right] &=&%
\sum_{k=0}^{2l-3}C_{2l-2}^{2l-2-k}\wp ^{\left( 2l-3-k\right) }\partial
^{k},\qquad \left[ \partial ^{2l-1-2j},\wp \right] =%
\sum_{k=0}^{2l-2-2j}C_{2l-1-2j}^{2l-1-2j-k}\wp ^{\left( 2l-1-2j-k\right)
}\partial ^{k}
\end{eqnarray*}%
and, hence,%
\begin{eqnarray*}
\begin{array}{c}
-\frac{1}{1+m}\mathcal{W}_{2l}%
\end{array}
&=&%
\begin{array}{c}
\frac{1}{2l\left( 2l-1\right) }\sum\limits_{k=1}^{2l-1}C_{2l-1}^{2l-k}\wp
^{\left( 2l-k\right) }\partial ^{k}+\frac{1}{2l\left( 2l-1\right) }%
\sum\limits_{k=0}^{2l-2}C_{2l-1}^{2l-1-k}\wp ^{\left( 2l-k\right) }\partial
^{k}%
\end{array}
\\
&&%
\begin{array}{c}
-\frac{1}{2}\wp ^{\prime \prime }\partial ^{2l-2}-\frac{1}{2l}\wp ^{\prime
}\partial ^{2l-1}-\sum\limits_{k=0}^{2l-3}C_{2l-2}^{2l-2-k}\wp ^{\prime }\wp
^{\left( 2l-3-k\right) }\partial
^{k}-\sum\limits_{k=0}^{2l-3}C_{2l-2}^{2l-2-k}\wp \wp ^{\left( 2l-2-k\right)
}\partial ^{k}%
\end{array}
\\
&&%
\begin{array}{c}
+\sum\limits_{j=2}^{l-1}\frac{\left( 2l-2\right) !}{\left( 2l-2j\right) !}%
\gamma _{2j-2}\sum\limits_{k=1}^{2l-1-2j}C_{2l-1-2j}^{2l-2j-k}\wp ^{\left(
2l-2j-k\right) }\partial ^{k}%
\end{array}
\\
&&%
\begin{array}{c}
+\sum\limits_{j=2}^{l-1}\frac{\left( 2l-2\right) !}{\left( 2l-2j\right) !}%
\gamma _{2j-2}\sum\limits_{k=0}^{2l-2-2j}C_{2l-1-2j}^{2l-1-2j-k}\wp ^{\left(
2l-2j-k\right) }\partial ^{k}+\sum\limits_{j=2}^{l-1}\left( 2j-2\right)
!C_{2l-1}^{2l-2j}\gamma _{2j-2}\wp ^{\prime }\partial ^{2l-2j-1}%
\end{array}%
\end{eqnarray*}%
which can be rewritten as%
\begin{eqnarray*}
\begin{array}{c}
-\frac{1}{1+m}\mathcal{W}_{2l}%
\end{array}
&=&%
\begin{array}{c}
\frac{1}{2l\left( 2l-1\right) }\sum\limits_{k=1}^{2l-1}C_{2l-1}^{2l-k}\wp
^{\left( 2l-k\right) }\partial ^{k}+\frac{1}{2l\left( 2l-1\right) }%
\sum\limits_{k=0}^{2l-2}C_{2l-1}^{2l-1-k}\wp ^{\left( 2l-k\right) }\partial
^{k}%
\end{array}
\\
&&%
\begin{array}{c}
-\frac{1}{2}\wp ^{\prime \prime }\partial ^{2l-2}-\frac{1}{2l}\wp ^{\prime
}\partial ^{2l-1}-\sum\limits_{k=0}^{2l-3}C_{2l-2}^{2l-2-k}\left( \wp
^{\prime }\wp ^{\left( 2l-3-k\right) }+\wp \wp ^{\left( 2l-2-k\right)
}\right) \partial ^{k}%
\end{array}
\\
&&%
\begin{array}{c}
+\sum\limits_{k=1}^{2l-5}\sum\limits_{j=2}^{l-1-\left[ \frac{k}{2}\right] }%
\frac{\left( 2l-2\right) !}{\left( 2l-2j\right) !}\gamma
_{2j-2}C_{2l-1-2j}^{2l-2j-k}\wp ^{\left( 2l-2j-k\right) }\partial ^{k}%
\end{array}
\\
&&%
\begin{array}{c}
+\sum\limits_{k=0}^{2l-6}\sum\limits_{j=2}^{l-1-\left[ \frac{k+1}{2}\right] }%
\frac{\left( 2l-2\right) !}{\left( 2l-2j\right) !}\gamma
_{2j-2}C_{2l-1-2j}^{2l-1-2j-k}\wp ^{\left( 2l-2j-k\right) }\partial
^{k}+\sum\limits_{k=1}^{l-2}\left( 2l-2k-2\right) !C_{2l-1}^{2k}\gamma
_{2l-2k-2}\wp ^{\prime }\partial ^{2k-1}.%
\end{array}%
\end{eqnarray*}%
It is shown below that the coefficients by all $\partial ^{k},$ $%
k=0,1,\ldots 2l-1,$ are zero. We have%
\begin{equation*}
\begin{array}{ll}
\partial ^{2l-1}: & \frac{1}{2l\left( 2l-1\right) }C_{2l-1}^{1}\wp ^{\prime
}-\frac{1}{2l}\wp ^{\prime }=0, \\ 
\partial ^{2l-2}: & \frac{1}{2l\left( 2l-1\right) }C_{2l-1}^{2}\wp ^{\prime
\prime }+\frac{1}{2l\left( 2l-1\right) }C_{2l-1}^{1}\wp ^{\prime \prime }-%
\frac{1}{2}\wp ^{\prime \prime }=0, \\ 
\partial ^{2l-3}: & 
\begin{array}{l}
\frac{1}{2l\left( 2l-1\right) }C_{2l-1}^{3}\wp ^{\left( 3\right) }+\frac{1}{%
2l\left( 2l-1\right) }C_{2l-1}^{2}\wp ^{\left( 3\right) }-2C_{2l-2}^{1}\wp
^{\prime }\wp \\ 
\qquad \qquad =\left( \frac{12}{2l\left( 2l-1\right) }C_{2l-1}^{3}+\frac{12}{%
2l\left( 2l-1\right) }C_{2l-1}^{2}-2C_{2l-2}^{1}\right) \wp ^{\prime }\wp =0,%
\end{array}
\\ 
\partial ^{2l-4}: & 
\begin{array}{l}
\frac{1}{2l\left( 2l-1\right) }C_{2l-1}^{4}\wp ^{\left( 4\right) }+\frac{1}{%
2l\left( 2l-1\right) }C_{2l-1}^{3}\wp ^{\left( 4\right) }-C_{2l-2}^{2}\left(
\wp ^{\prime }\wp ^{\prime }+\wp \wp ^{\left( 2\right) }\right) \\ 
\qquad \qquad =\left( \frac{12}{2l\left( 2l-1\right) }C_{2l-1}^{4}+\frac{12}{%
2l\left( 2l-1\right) }C_{2l-1}^{3}-C_{2l-2}^{2}\right) \left( \wp ^{\prime
}\wp ^{\prime }+\wp \wp ^{\left( 2\right) }\right) =0.%
\end{array}%
\end{array}%
\end{equation*}%
When $k=2l-2q-1,$ $q=2,...,l-1,$ coefficient by $\partial ^{2l-2q-1}$ is%
\begin{equation*}
\begin{array}{c}
K_{2l-2q-1}=\frac{\left( 2l-2\right) !}{\left( 2l-2q-1\right) !}\left( \frac{%
\wp ^{\left( 2q+1\right) }}{\left( 2q+1\right) !}-\frac{\wp ^{\prime }\wp
^{\left( 2q-2\right) }+\wp \wp ^{\left( 2q-1\right) }}{\left( 2q-1\right) !}%
+\sum\limits_{j=2}^{q-1}\frac{\gamma _{2j-2}}{\left( 2q+1-2j\right) !}\wp
^{\left( 2q+1-2j\right) }+\frac{2q}{2q-1}\gamma _{2q-2}\wp ^{\prime }\right)
.%
\end{array}%
\end{equation*}%
To show that this coefficient is zero let us calculate its Laurent expansion
and show that it consists of the terms by the positive degrees of $z$ only.
Since it is also a doubly periodic function it can only be zero. The Laurent
expansion for the derivatives of the Weierstrass function are given by%
\begin{eqnarray*}
\wp ^{\prime } &=&-2z^{-3}+O\left( z\right) , \\
\wp ^{\left( 2q+1\right) } &=&-\left( 2q+2\right) !z^{-\left( 2q+3\right)
}+O\left( z\right) , \\
\wp ^{\left( 2q+1-2j\right) } &=&-\left( 2q+2-2j\right) !z^{-\left(
2q-2j+3\right) }+O\left( z\right) .
\end{eqnarray*}%
Therefore, we can find that%
\begin{eqnarray*}
\wp \wp ^{\left( 2q-1\right) } &=&-\left( 2q\right) !z^{-\left( 2q+3\right)
}-\left( 2q\right) !\sum_{i=1}^{q-1}\gamma _{2i}z^{-\left( 2q+1-2i\right)
}+O\left( z\right) , \\
\wp ^{\prime }\wp ^{\left( 2q-2\right) } &=&-2\left( 2q-1\right) !z^{-\left(
2q+3\right) }-2\left( 2q-2\right) !\gamma _{2q-2}z^{-3}+\left( 2q-1\right)
!\sum_{i=1}^{q-1}2i\gamma _{2i}z^{-\left( 2q+1-2i\right) }+O\left( z\right)
\end{eqnarray*}%
and%
\begin{equation*}
\begin{array}{c}
-\frac{\left( \wp ^{\prime }\wp ^{\left( 2q-2\right) }+\wp \wp ^{\left(
2q-1\right) }\right) }{\left( 2q-1\right) !}=\left( 2q+2\right) z^{-\left(
2q+3\right) }+\sum\limits_{i=1}^{q-2}\left( 2q-2i\right) \gamma
_{2i}z^{-\left( 2q+1-2i\right) }+\frac{4q}{2q-1}\gamma _{2q-2}z^{-3}+O\left(
z\right) .%
\end{array}%
\end{equation*}%
Therefore%
\begin{eqnarray*}
K_{2l-2q-1} &=&%
\begin{array}{c}
\frac{\left( 2l-2\right) !}{\left( 2l-2q-1\right) !}\left( 
\rule[12pt]{0pt}{12pt}-\left( 2q+2\right) z^{-\left( 2q+3\right) }+\left(
2q+2\right) z^{-\left( 2q+3\right) }+\sum_{i=1}^{q-2}\left( 2q-2i\right)
\gamma _{2i}z^{-\left( 2q+1-2i\right) }\right.%
\end{array}
\\
&&%
\begin{array}{c}
\qquad \qquad \qquad +\left. \frac{4q}{2q-1}\gamma
_{2q-2}z^{-3}-\sum_{j=2}^{q-1}\gamma _{2j-2}\left( 2q+2-2j\right) z^{-\left(
2q-2j+3\right) }-\gamma _{2q-2}\frac{4q}{2q-1}z^{-3}+O\left( z\right) 
\rule[12pt]{0pt}{12pt}\right)%
\end{array}
\\
&=&%
\begin{array}{c}
\frac{\left( 2l-2\right) !}{\left( 2l-2q-1\right) !}\left( O\left( z\right)
\right) =0.%
\end{array}%
\end{eqnarray*}%
When $k=2l-2q;q=3,...,l-1,$ coefficient by $\partial ^{2l-2q}$ is%
\begin{equation*}
\begin{array}{c}
K_{2l-2q}=\frac{\left( 2l-2\right) !}{\left( 2l-2q\right) !}\left( \frac{1}{%
\left( 2q\right) !}\wp ^{\left( 2q\right) }-\frac{\left( \wp ^{\prime }\wp
^{\left( 2q-3\right) }+\wp \wp ^{\left( 2q-2\right) }\right) }{\left(
2q-2\right) !}+\sum_{j=2}^{q-1}\frac{1}{\left( 2q-2j\right) !}\gamma
_{2j-2}\wp ^{\left( 2q-2j\right) }\right) .%
\end{array}%
\end{equation*}%
The Laurent expansion for the derivatives of the Weierstrass function are
given by%
\begin{eqnarray*}
\wp ^{\left( 2q\right) } &=&\left( 2q+1\right) !z^{-\left( 2q+2\right)
}+\left( 2q\right) !\gamma _{2q}+O\left( z^{2}\right) , \\
\wp ^{\left( 2q-2j\right) } &=&\left( 2q-2j+1\right) !z^{-\left(
2q-2j+2\right) }+\left( 2q-2j\right) !\gamma _{2q-2j}+O\left( z^{2}\right)
\end{eqnarray*}%
and we calculate that%
\begin{eqnarray*}
\wp \wp ^{\left( 2q-2\right) } &=&\left( 2q-1\right) !z^{-\left( 2q+2\right)
}+\left( 2q-2\right) !2q\gamma _{2q-2}z^{-2}+\left( 2q-1\right) !\left(
q+1\right) \gamma _{2q} \\
&&+\left( 2q-1\right) !\left( \sum_{i=1}^{q-2}\gamma _{2i}z^{-\left(
2q-2i\right) }\right) +O\left( z^{2}\right) , \\
\wp ^{\prime }\wp ^{\left( 2q-3\right) } &=&2\left( 2q-2\right) !z^{-\left(
2q+2\right) }-\left( 2q-2\right) !2q\gamma _{2q-2}z^{-2}-\left( 2q-2\right)
!\left( 2q\right) \left( \frac{2q+2}{3}\right) \gamma _{2q} \\
&&-\left( 2q-2\right) !\sum_{i=1}^{q-2}2i\gamma _{2i}z^{-\left( 2q-2i\right)
}+O\left( z^{2}\right)
\end{eqnarray*}%
and%
\begin{equation*}
\begin{array}{c}
-\frac{\left( \wp \wp ^{\left( 2q-2\right) }+\wp ^{\prime }\wp ^{\left(
2q-3\right) }\right) }{\left( 2q-2\right) !}=-\left( 2q+1\right) z^{-\left(
2q+2\right) }-\sum\limits_{i=1}^{q-2}\left( 2q-1-2i\right) \gamma
_{2i}z^{-\left( 2q-2i\right) }-\left( \frac{\left( 1+q\right) \left(
2q-3\right) }{3}\right) \gamma _{2q}+O\left( z^{2}\right) .%
\end{array}%
\end{equation*}%
Therefore%
\begin{eqnarray*}
K_{2l-2q} &=&%
\begin{array}{c}
\frac{\left( 2l-2\right) !}{\left( 2l-2q\right) !}\left( 
\rule[12pt]{0pt}{12pt}\left( 2q+1\right) z^{-\left( 2q+2\right) }+\gamma
_{2q}-\left( 2q+1\right) z^{-\left( 2q+2\right) }-\sum_{i=1}^{q-2}\left(
2q-1-2i\right) \gamma _{2i}z^{-\left( 2q-2i\right) }\right.%
\end{array}
\\
&&\qquad \qquad 
\begin{array}{c}
\left. -\left( \frac{\left( 1+q\right) \left( 2q-3\right) }{3}\right) \gamma
_{2q}+\sum_{j=2}^{q-1}\gamma _{2j-2}\left( 2q-2j+1\right) z^{-\left(
2q-2j+2\right) }+\sum\limits_{j=2}^{q-1}\gamma _{2j-2}\gamma
_{2q-2j}+O\left( z^{2}\right) \rule[12pt]{0pt}{12pt}\right)%
\end{array}
\\
&=&%
\begin{array}{c}
\frac{\left( 2l-2\right) !}{\left( 2l-2q\right) !}\left( -\frac{\left(
2q+3\right) \left( q-2\right) }{3}\gamma _{2q}+\sum\limits_{j=2}^{q-1}\gamma
_{2j-2}\gamma _{2q-2j}+O\left( z^{2}\right) \right) =0.%
\end{array}%
\end{eqnarray*}%
Here we have used the identity (\ref{gammas_relationship}).

The last case to consider is when $k=0.$ In this case we need to show that%
\begin{equation*}
\begin{array}{l}
K_{0}=\frac{1}{2l\left( 2l-1\right) }\wp ^{\left( 2l\right) }-\left( \wp
^{\prime }\wp ^{\left( 2l-3\right) }+\wp \wp ^{\left( 2l-2\right) }\right)
+\sum\limits_{j=2}^{l-1}\frac{\left( 2l-2\right) !}{\left( 2l-2j\right) !}%
\gamma _{2j-2}\wp ^{\left( 2l-2j\right) }%
\end{array}%
\end{equation*}%
is zero. We have%
\begin{eqnarray*}
K_{0} &=&%
\begin{array}{c}
\left( 2l-2\right) !\left( \rule[12pt]{0pt}{12pt}\left( 2l+1\right)
z^{-\left( 2l+2\right) }+\gamma _{2l}-\left( 2l+1\right) z^{-\left(
2l+2\right) }-\sum\limits_{i=1}^{l-2}\left( 2l-1-2i\right) \gamma
_{2i}z^{-\left( 2l-2i\right) }\right.%
\end{array}
\\
&&%
\begin{array}{c}
\qquad \qquad \qquad \left. -\left( \frac{\left( 1+l\right) \left(
2l-3\right) }{3}\right) \gamma _{2l}+\sum\limits_{j=2}^{l-1}\gamma
_{2j-2}\left( 2l-2j+1\right) z^{-\left( 2l-2j+2\right)
}+\sum\limits_{j=2}^{l-1}\gamma _{2j-2}\gamma _{2l-2j}+O\left( z^{2}\right) 
\rule[12pt]{0pt}{12pt}\right)%
\end{array}
\\
&=&%
\begin{array}{c}
\left( 2l-2\right) !\left( -\frac{\left( 2l+3\right) \left( l-2\right) }{3}%
\gamma _{2l}+\sum_{j=2}^{l-1}\gamma _{2j-2}\gamma _{2l-2j}+O\left(
z^{2}\right) \right) =0.%
\end{array}%
\end{eqnarray*}%
The proof of Lemma 3 is now finished.

\textbf{Lemma 4. }The following identity holds%
\begin{equation*}
\begin{array}{c}
\left[ \Theta ,\sum\limits_{j=2}^{n}u_{1j}\right] +m\sum\limits_{j=2}^{n}%
\left[ \Theta _{\hat{\jmath}},u_{1j}\right] \partial -\frac{\left(
1+m\right) }{2}\sum\limits_{j=2}^{n}u_{1j}^{\prime }\Theta _{\hat{\jmath}}-%
\frac{\left( 1-m\right) }{2}\sum\limits_{j=2}^{n}\Theta _{\hat{\jmath}%
}u_{1j}^{\prime }+m\sum\limits_{k=2}^{n}\sum\limits_{\substack{ l=2  \\ %
l\neq k}}^{n}\left[ \Theta _{\hat{k}\hat{l}},u_{1k}\right] u_{1l}=0.%
\end{array}%
\end{equation*}%
Here again $\partial =\frac{\partial }{\partial x_{1}}.$

\textbf{Proof. }From Lemma 3 
\begin{equation}
\begin{array}{c}
\left[ \mathfrak{D}^{n},u_{1j}\right] +\left[ \left[ \varsigma _{j},%
\mathfrak{D}^{n-1}\right] ,u_{1j}\right] +m\left[ \mathfrak{D}^{n-1},u_{1j}%
\right] \partial -\frac{\left( 1+m\right) }{2}u_{1j}^{\prime }\mathfrak{D}%
^{n-1}-\frac{\left( 1-m\right) }{2}\mathfrak{D}^{n-1}u_{1j}^{\prime }=0.%
\end{array}
\label{D_u}
\end{equation}%
Note also that%
\begin{eqnarray*}
\begin{array}{c}
\sum\limits_{j=2}^{n}\sum\limits_{\sigma \in \mathfrak{S}\left( \hat{1}\hat{%
\jmath};t\right) }ad_{\varsigma _{\sigma }}^{t}\left( \left[ \mathfrak{D}%
^{n-t-1},u_{1j}\right] \partial \right)%
\end{array}
&=&%
\begin{array}{c}
\sum\limits_{j=2}^{n}\sum\limits_{\sigma \in \mathfrak{S}\left( \hat{1}\hat{%
\jmath};t\right) }ad_{\varsigma _{\sigma }}^{t}\left( \left[ \mathfrak{D}%
^{n-t-1},u_{1j}\right] \right) \partial%
\end{array}
\\
&&%
\begin{array}{c}
+\sum\limits_{k=2}^{n}\sum\limits_{\substack{ l=2  \\ l\neq k}}%
^{n}\sum_{\sigma \in \mathfrak{S}\left( \hat{1}\hat{k}\hat{l};t\right) }%
\left[ ad_{\varsigma _{\sigma }}^{t}\left( \mathfrak{D}^{n-2-t}\right)
,u_{1k}\right] u_{1l}.%
\end{array}%
\end{eqnarray*}%
We sum (\ref{D_u}) and then use (\ref{Theta}) to obtain $\Theta $ and $%
\Theta _{\hat{k}}^{n}.$ We have%
\begin{eqnarray*}
0 &=&%
\begin{array}{c}
\sum\limits_{t=0}^{\left[ \frac{n}{2}\right] }\sum\limits_{j=2}^{n}\sum%
\limits_{\sigma \in \mathfrak{S}\left( \hat{1}\hat{\jmath};t\right)
}ad_{\varsigma _{\sigma }}^{t}\left( \rule[6pt]{0pt}{12pt}\left[ \mathfrak{D}%
^{n-t},u_{1j}\right] +\left[ \left[ \varsigma _{j},\mathfrak{D}^{n-t-1}%
\right] ,u_{1j}\right] +m\left[ \mathfrak{D}^{n-t-1},u_{1j}\right] \partial
\right.%
\end{array}
\\
&&%
\begin{array}{c}
\qquad \qquad \qquad \qquad \qquad \qquad \qquad \qquad \qquad \qquad \qquad
\left. -\frac{\left( 1+m\right) }{2}u_{1j}^{\prime }\mathfrak{D}^{n-t-1}-%
\frac{\left( 1-m\right) }{2}\mathfrak{D}^{n-t-1}u_{1j}^{\prime }%
\rule[6pt]{0pt}{12pt}\right)%
\end{array}
\\
&=&%
\begin{array}{c}
\sum\limits_{t=0}^{\left[ \frac{n}{2}\right] }\sum\limits_{j=2}^{n}\sum%
\limits_{\sigma \in \mathfrak{S}\left( \hat{1}\hat{\jmath};t\right) }\left[
ad_{\varsigma _{\sigma }}^{t}\left( \mathfrak{D}^{n-t}\right) ,u_{1j}\right]
+\sum\limits_{t=0}^{\left[ \frac{n-1}{2}\right] }\sum\limits_{j=2}^{n}\sum%
\limits_{\sigma \in \mathfrak{S}\left( \hat{1}\hat{\jmath};t\right) }\left[
ad_{\varsigma _{\left\{ j,\sigma \right\} }}^{t}\left( \mathfrak{D}%
^{n-t-1}\right) ,u_{1j}\right]%
\end{array}
\\
&&%
\begin{array}{c}
+m\sum\limits_{t=0}^{\left[ \frac{n-1}{2}\right] }\sum\limits_{j=2}^{n}\sum%
\limits_{\sigma \in \mathfrak{S}\left( \hat{1}\hat{\jmath};t\right)
}ad_{\varsigma _{\sigma }}^{t}\left( \left[ \mathfrak{D}^{n-t-1},u_{1j}%
\right] \right) \partial +m\sum\limits_{t=0}^{\left[ \frac{n-2}{2}\right]
}\sum\limits_{k=2}^{n}\sum\limits_{\substack{ l=2  \\ l\neq k}}%
^{n}\sum\limits_{\sigma \in \mathfrak{S}\left( \hat{1}\hat{k}\hat{l}%
;t\right) }\left[ ad_{\varsigma _{\sigma }}^{t}\left( \mathfrak{D}%
^{n-2-t}\right) ,u_{1k}\right] u_{1l}%
\end{array}
\\
&&%
\begin{array}{c}
-\frac{\left( 1+m\right) }{2}\sum\limits_{t=0}^{\left[ \frac{n-1}{2}\right]
}\sum\limits_{j=2}^{n}\sum\limits_{\sigma \in \mathfrak{S}\left( \hat{1}\hat{%
\jmath};t\right) }u_{1j}^{\prime }ad_{\varsigma _{\sigma }}^{t}\left( 
\mathfrak{D}^{n-t-1}\right) -\frac{\left( 1-m\right) }{2}\sum\limits_{t=0}^{%
\left[ \frac{n-1}{2}\right] }\sum\limits_{j=2}^{n}\sum\limits_{\sigma \in 
\mathfrak{S}\left( \hat{1}\hat{\jmath};t\right) }ad_{\varsigma _{\sigma
}}^{t}\left( \mathfrak{D}^{n-t-1}\right) u_{1j}^{\prime }%
\end{array}%
\end{eqnarray*}%
\begin{eqnarray*}
&=&%
\begin{array}{c}
\sum\limits_{t=0}^{\left[ \frac{n}{2}\right] }\sum\limits_{\sigma \in 
\mathfrak{S}\left( \hat{1};t\right) }\left[ ad_{\varsigma _{\sigma
}}^{t}\left( \mathfrak{D}^{n-t}\right) ,\sum_{j=2}^{n}u_{1j}\right]
+m\sum\limits_{t=0}^{\left[ \frac{n-1}{2}\right] }\sum\limits_{j=2}^{n}\sum%
\limits_{\sigma \in \mathfrak{S}\left( \hat{1}\hat{\jmath};t\right)
}ad_{\varsigma _{\sigma }}^{t}\left( \left[ \mathfrak{D}^{n-t-1},u_{1j}%
\right] \right) \partial%
\end{array}
\\
&&%
\begin{array}{c}
+m\sum\limits_{t=0}^{\left[ \frac{n-2}{2}\right] }\sum\limits_{k=2}^{n}\sum%
\limits_{\substack{ l=2  \\ l\neq k}}^{n}\sum\limits_{\sigma \in \mathfrak{S}%
\left( \hat{1}\hat{k}\hat{l};t\right) }\left[ ad_{\varsigma _{\sigma
}}^{t}\left( \mathfrak{D}^{n-2-t}\right) ,u_{1k}\right] u_{1l}%
\end{array}
\\
&&%
\begin{array}{c}
-\frac{\left( 1+m\right) }{2}\sum\limits_{t=0}^{\left[ \frac{n-1}{2}\right]
}\sum\limits_{j=2}^{n}\sum\limits_{\sigma \in \mathfrak{S}\left( \hat{1}\hat{%
\jmath};t\right) }u_{1j}^{\prime }ad_{\varsigma _{\sigma }^{\mu }}^{t}\left( 
\mathfrak{D}^{n-t-1}\right) -\frac{\left( 1-m\right) }{2}\sum\limits_{t=0}^{%
\left[ \frac{n-1}{2}\right] }\sum\limits_{j=2}^{n}\sum\limits_{\sigma \in 
\mathfrak{S}\left( \hat{1}\hat{\jmath};t\right) }ad_{\varsigma _{\sigma
}}^{t}\left( \mathfrak{D}^{n-t-1}\right) u_{1j}^{\prime }%
\end{array}
\\
&=&%
\begin{array}{c}
\left[ \Theta ,\sum\limits_{j=2}^{n}u_{1j}\right] +m\sum\limits_{j=2}^{n}%
\left[ \Theta _{\hat{\jmath}},u_{1j}\right] \partial -\frac{\left(
1+m\right) }{2}\sum\limits_{j=2}^{n}u_{1j}^{\prime }\Theta _{\hat{\jmath}}-%
\frac{\left( 1-m\right) }{2}\sum\limits_{j=2}^{n}\Theta _{\hat{\jmath}%
}u_{1j}^{\prime }+m\sum\limits_{k=2}^{n}\sum\limits_{\substack{ l=2  \\ %
l\neq k}}^{n}\left[ \Theta _{\hat{k}\hat{l}},u_{1k}\right] u_{1l}.%
\end{array}%
\end{eqnarray*}%
Lemma 4 is proven.

\textbf{Lemma 5. }The following identity holds%
\begin{equation}
\left[ \Delta ,\Theta \right] =2\sum\limits_{k=2}^{n}\left[ u_{1k},\Theta _{%
\hat{k}}\right] \partial _{k}-2m\left( \sum\limits_{k=2}^{n}\left[
u_{1k},\Theta _{\hat{k}}\right] \right) \partial _{1}-\left( 1+m\right)
\sum\limits_{k=2}^{n}\left[ u_{1k}^{\prime },\Theta _{\hat{k}}\right]
+m\sum\limits_{k=2}^{n}\sum\limits_{\substack{ l=2  \\ l\neq k}}^{n}\left[
u_{1k},\left[ u_{1l},\Theta _{\hat{k}\hat{l}}\right] \right] .
\label{Delta,Theta}
\end{equation}

\textbf{Proof.}%
\begin{equation*}
\left[ \Delta ,\Theta \right] =2m\frac{\partial \Theta }{\partial x_{1}}%
\partial _{1}+2\sum\limits_{k=2}^{n}\frac{\partial \Theta }{\partial x_{k}}%
\partial _{k}+m\frac{\partial ^{2}\Theta }{\partial x_{1}^{2}}%
+\sum\limits_{k=2}^{n}\frac{\partial ^{2}\Theta }{\partial x_{k}^{2}}.
\end{equation*}%
From (\ref{Theta}) it follows that $\frac{\partial \Theta }{\partial x_{k}}=%
\left[ u_{1k},\Theta _{\hat{k}}\right] $ and $\frac{\partial \Theta }{%
\partial x_{1}}=-\sum\limits_{k=2}^{n}\left[ u_{1k},\Theta _{\hat{k}}\right]
,$ therefore $\frac{\partial ^{2}\Theta }{\partial x_{k}^{2}}=-\left[
u_{1k}^{\prime },\Theta _{\hat{k}}\right] $ and $\frac{\partial ^{2}\Theta }{%
\partial x_{1}^{2}}=-\sum\limits_{k=2}^{n}\left[ u_{1k}^{\prime },\Theta _{%
\hat{k}}\right] +\sum\limits_{k=2}^{n}\sum\limits_{\substack{ l=2  \\ l\neq
k }}^{n}\left[ u_{1k},\left[ u_{1l},\Theta _{\hat{k}\hat{l}}\right] \right]
, $and we arrive at (\ref{Delta,Theta}).

\textbf{Lemma 6.} 
\begin{equation*}
R=\left[ \Theta ,H\right] +\sum\limits_{j=2}^{n}\left[ \Theta _{\hat{\jmath}%
}\partial _{j},2\sum\limits_{\substack{ l=1  \\ l\neq j}}^{n}u_{jl}\right]
+\sum\limits_{2\leq k<l\leq n}\left[ \Theta _{\hat{k}\hat{l}},2\left(
u_{1k}+u_{1l}\right) \right] X_{\left\{ k,l\right\} }=0.
\end{equation*}

\textbf{Proof.} We have%
\begin{equation*}
R=\left[ \Delta ,\Theta \right] +2\sum\limits_{k=2}^{n}\left[ \Theta ,u_{1k}%
\right] +2\sum\limits_{j=2}^{n}\sum\limits_{\substack{ l=1  \\ l\neq j}}^{n}%
\left[ \Theta _{\hat{\jmath}},u_{jl}\right] \partial
_{j}+2\sum\limits_{j=2}^{n}\sum\limits_{\substack{ l=1  \\ l\neq j}}%
^{n}\Theta _{\hat{\jmath}}\left[ \partial _{j},u_{jl}\right]
+2\sum\limits_{2\leq k<l\leq n}\left[ \Theta _{\hat{k}\hat{l}},u_{1k}+u_{1l}%
\right] u_{kl}
\end{equation*}%
Using Lemma 5 $R$ is rewritten as%
\begin{eqnarray*}
R &=&2\sum\limits_{k=2}^{n}\left[ u_{1k},\Theta _{\hat{k}}\right] \partial
_{k}-2m\sum\limits_{k=2}^{n}\left[ u_{1k},\Theta _{\hat{k}}\right] \partial
_{1}-\left( 1+m\right) \sum\limits_{k=2}^{n}\left[ u_{1k}^{\prime },\Theta _{%
\hat{k}}\right] +m\sum\limits_{k=2}^{n}\sum\limits_{\substack{ l=2  \\ l\neq
k }}^{n}\left[ u_{1k},\left[ u_{1l},\Theta _{\hat{k}\hat{l}}\right] \right]
\\
&&+2\sum\limits_{k=2}^{n}\left[ \Theta ,u_{1k}\right] +2\sum\limits_{j=2}^{n}%
\left[ \Theta _{\hat{\jmath}},u_{1j}\right] \partial
_{j}+2\sum\limits_{j=2}^{n}\sum\limits_{\substack{ l=1  \\ l\neq j}}%
^{n}\Theta _{\hat{\jmath}}u_{jl}^{\prime }+2\sum\limits_{2\leq k<l\leq n}%
\left[ \Theta _{\hat{k}\hat{l}},u_{1k}+u_{1l}\right] u_{kl}
\end{eqnarray*}%
and, hence,%
\begin{eqnarray*}
R &=&-2m\sum\limits_{k=2}^{n}\left[ u_{1k},\Theta _{\hat{k}}\right] \partial
-\left( 1+m\right) \sum\limits_{k=2}^{n}\left[ u_{1k}^{\prime },\Theta _{%
\hat{k}}\right] +m\sum\limits_{k=2}^{n}\sum\limits_{\substack{ l=2  \\ l\neq
k }}^{n}\left[ u_{1k},\left[ u_{1l},\Theta _{\hat{k}\hat{l}}\right] \right]
+2\sum\limits_{k=2}^{n}\left[ \Theta ,u_{1k}\right] \\
&&-2\sum\limits_{j=2}^{n}\Theta _{\hat{\jmath}}u_{1j}^{\prime
}+2\sum\limits_{2\leq k<l\leq n}\left( \Theta _{\hat{k}}-\Theta _{\hat{l}%
}\right) u_{kl}^{\prime }+2\sum\limits_{2\leq k<l\leq n}\left[ \Theta _{\hat{%
k}\hat{l}},u_{1k}+u_{1l}\right] u_{kl}.
\end{eqnarray*}%
Now, let us apply Lemma 4 which simplifies the above expression into the
following%
\begin{eqnarray*}
R &=&\left( 1+m\right) \sum\limits_{k=2}^{n}u_{1k}^{\prime }\Theta _{\hat{k}%
}+\left( 1-m\right) \sum\limits_{k=2}^{n}\Theta _{\hat{k}}u_{1k}^{\prime
}-2m\sum\limits_{k=2}^{n}\sum\limits_{\substack{ l=2  \\ l\neq k}}^{n}\left[
\Theta _{\hat{k}\hat{l}},u_{1k}\right] u_{1l}-\left( 1+m\right)
\sum\limits_{k=2}^{n}u_{1k}^{\prime }\Theta _{\hat{k}} \\
&&+\left( 1+m\right) \sum\limits_{k=2}^{n}\Theta _{\hat{k}}u_{1k}^{\prime
}+m\sum\limits_{k=2}^{n}\sum\limits_{\substack{ l=2  \\ l\neq k}}^{n}\left[
u_{1k},\left[ u_{1l},\Theta _{\hat{k}\hat{l}}\right] \right]
+\sum\limits_{2\leq k<l\leq n}\left[ \Theta _{\hat{k}\hat{l}},2\left(
u_{1k}+u_{1l}\right) \right] u_{kl} \\
&&-2\sum\limits_{j=2}^{n}\Theta _{\hat{\jmath}}u_{1j}^{\prime
}+2\sum\limits_{2\leq k<l\leq n}\left( \Theta _{\hat{k}}-\Theta _{\hat{l}%
}\right) u_{kl}^{\prime } \\
&=&-2m\sum\limits_{k=2}^{n}\sum\limits_{\substack{ l=2  \\ l\neq k}}^{n}%
\left[ \Theta _{\hat{k}\hat{l}},u_{1k}\right] u_{1l}+m\sum\limits_{k=2}^{n}%
\sum\limits_{\substack{ l=2  \\ l\neq k}}^{n}\left[ u_{1k},\left[
u_{1l},\Theta _{\hat{k}\hat{l}}\right] \right] \\
&&+2\sum\limits_{2\leq k<l\leq n}\left[ \Theta _{\hat{k}\hat{l}},\left(
u_{1k}+u_{1l}\right) \right] u_{kl}+2\sum\limits_{2\leq k<l\leq n}\left(
\Theta _{\hat{k}}-\Theta _{\hat{l}}\right) u_{kl}^{\prime }.
\end{eqnarray*}%
From (\ref{Theta}) $\Theta _{\hat{k}}-\Theta _{\hat{l}}=\left[ \varsigma
_{l}-\varsigma _{k},\Theta _{\hat{k}\hat{l}}\right] $ and, therefore, $R$ is
simplified to%
\begin{eqnarray*}
R &=&m\sum\limits_{k=2}^{n}\sum\limits_{\substack{ l=2  \\ l\neq k}}^{n}%
\left[ u_{1k}u_{1l},\Theta _{\hat{k}\hat{l}}\right] +2\sum\limits_{2\leq
k<l\leq n}\left[ \Theta _{\hat{k}\hat{l}},\left( u_{1k}+u_{1l}\right) u_{kl}%
\right] +2\sum\limits_{2\leq k<l\leq n}\left[ \varsigma _{1l}-\varsigma
_{1k},\Theta _{\hat{k}\hat{l}}\right] u_{kl}^{\prime } \\
&=&2\sum\limits_{2\leq k<l\leq n}\left[ \left( \varsigma _{l}-\varsigma
_{k}\right) u_{kl}^{\prime }-\left( u_{1k}+u_{1l}\right)
u_{kl}+mu_{1k}u_{1l},\Theta _{\hat{k}\hat{l}}\right] .
\end{eqnarray*}%
All derivatives with respect to $x_{1}$ of the term 
\begin{multline*}
\left( \varsigma _{l}-\varsigma _{k}\right) u_{kl}^{\prime }-\left(
u_{1k}+u_{1l}\right) u_{kl}+mu_{1k}u_{1l} \\
=m\left( m+1\right) ^{2}\left( \rule[6pt]{0pt}{12pt}\left( 
\rule[3pt]{0pt}{12pt}\varsigma \left( x_{1}-x_{l}\right) -\varsigma \left(
x_{1}-x_{k}\right) \right) \wp _{kl}^{\prime }-\left( \wp _{1k}+\wp
_{1l}\right) \wp _{kl}+\wp _{1k}\wp _{1l}\right)
\end{multline*}%
are zero. Indeed, this follows from the fact that the first derivative is
the addition theorem for the Weierstrass $\wp $-function. 
\begin{multline*}
\frac{\partial }{\partial x_{1}}\left( \rule[6pt]{0pt}{12pt}\left( 
\rule[3pt]{0pt}{12pt}\varsigma \left( x_{1}-x_{l}\right) -\varsigma \left(
x_{1}-x_{k}\right) \right) \wp _{kl}^{\prime }-\left( \wp _{1k}+\wp
_{1l}\right) \wp _{kl}+\wp _{1k}\wp _{1l}\right) \\
=\left( \wp _{1k}-\wp _{1l}\right) \wp _{kl}^{\prime }-\left( \wp
_{1k}^{\prime }+\wp _{1l}^{\prime }\right) \wp _{kl}+\wp _{1k}^{\prime }\wp
_{1l}+\wp _{1k}\wp _{1l}^{\prime }=0.
\end{multline*}%
Therefore $R=2\sum\nolimits_{2\leq k<l\leq n}\left[ \left( \varsigma
_{1l}-\varsigma _{1k}\right) u_{kl}^{\prime }-\left( u_{1k}+u_{1l}\right)
u_{kl}+mu_{1k}u_{1l},\Theta _{\hat{k}\hat{l}}\right] =0$ and Lemma 6 is
proven.

\textbf{Proof of Theorem 1.}

We use Lemma 3 to reduce the commutator $\left[ I,H\right] $ to the
expression (\ref{tails}) and we show below that (\ref{tails}) is zero. From (%
\ref{X})%
\begin{equation*}
X=\Theta +\sum\limits_{t=1}^{\left[ \frac{n-2}{2}\right] }\sum\limits_{%
\sigma \in \mathfrak{S}\left( \hat{1};2t\right) }X_{\sigma }\Theta _{\hat{%
\sigma}},
\end{equation*}%
therefore we can rewrite (\ref{tails}) as following%
\begin{eqnarray*}
\left[ I,H\right] &=&\left[ X_{\hat{1}}\partial
_{1},2\sum\nolimits_{l=2}^{n}u_{1l}\right] +\left[ \sum\limits_{k=2}^{n}X_{%
\hat{1}\hat{k}}\partial _{1}\partial _{k},2u_{1k}\right] +\left[ \Theta
+\sum\limits_{t=1}^{\left[ \frac{n-2}{2}\right] }\sum\limits_{\sigma \in 
\mathfrak{S}\left( \hat{1};2t\right) }X_{\sigma }\Theta _{\hat{\sigma}},H%
\right] \\
&&+\left[ \sum\nolimits_{j=2}^{n}\left( \Theta _{\hat{\jmath}%
}+\sum\limits_{t=1}^{\left[ \frac{n-3}{2}\right] }\sum\limits_{\sigma \in 
\mathfrak{S}\left( \hat{1}\hat{\jmath};2t\right) }X_{\sigma }\Theta _{\hat{%
\jmath}\hat{\sigma}}\right) \partial _{j},2\sum\limits_{\substack{ l=1  \\ %
l\neq j}}^{n}u_{jl}\right] \\
&&+\left[ \sum\limits_{2\leq i<j\leq n}\left( \Theta _{\hat{\imath}\hat{%
\jmath}}+\sum\limits_{t=1}^{\left[ \frac{n-4}{2}\right] }\sum\limits_{\sigma
\in \mathfrak{S}\left( \hat{1}\hat{\imath}\hat{\jmath};2t\right) }X_{\sigma
}\Theta _{\hat{\imath}\hat{\jmath}\hat{\sigma}}\right) \partial _{i}\partial
_{j},2u_{ij}\right]
\end{eqnarray*}%
\begin{eqnarray*}
&=&\left[ \Theta ,H\right] +\sum\limits_{t=1}^{\left[ \frac{n-2}{2}\right]
}\sum\limits_{\sigma \in \mathfrak{S}\left( \hat{1};2t\right) }\left[
X_{\sigma },-\Delta _{\sigma }\right] \Theta _{\hat{\sigma}%
}+\sum\limits_{t=1}^{\left[ \frac{n-2}{2}\right] }\sum\limits_{\sigma \in 
\mathfrak{S}\left( \hat{1};2t\right) }\left[ \Theta _{\hat{\sigma}},-\Delta
_{\hat{\sigma}}+2\sum\limits_{j=2}^{n}u_{1j}\right] X_{\sigma } \\
&&+\sum\nolimits_{j=2}^{n}\left[ \Theta _{\hat{\jmath}}\partial
_{j},2\sum\limits_{\substack{ l=1  \\ l\neq j}}^{n}u_{jl}\right]
+\sum\nolimits_{j=2}^{n}\sum\limits_{t=1}^{\left[ \frac{n-3}{2}\right]
}\sum\limits_{\sigma \in \mathfrak{S}\left( \hat{1}\hat{\jmath};2t\right) }%
\left[ X_{\sigma }\partial _{j},2\sum\limits_{l\in \sigma }u_{jl}\right]
\Theta _{\hat{\jmath}\hat{\sigma}} \\
&&+\sum\nolimits_{j=2}^{n}\sum\limits_{t=1}^{\left[ \frac{n-3}{2}\right]
}\sum\limits_{\sigma \in \mathfrak{S}\left( \hat{1}\hat{\jmath};2t\right) }%
\left[ \Theta _{\hat{\jmath}\hat{\sigma}}\partial _{j},2\sum\limits 
_{\substack{ l=1  \\ l\notin j\cup \sigma }}^{n}u_{jl}\right] X_{\sigma }+%
\left[ \sum\limits_{2\leq i<j\leq n}\partial _{i}\partial _{j},2u_{ij}\right]
\Theta _{\hat{\imath}\hat{\jmath}} \\
&&+\sum\limits_{2\leq i<j\leq n}\sum\limits_{t=1}^{\left[ \frac{n-4}{2}%
\right] }\sum\limits_{\sigma \in \mathfrak{S}\left( \hat{1}\hat{\imath}\hat{%
\jmath};2t\right) }\left[ X_{\sigma }\partial _{i}\partial _{j},2u_{ij}%
\right] \Theta _{\hat{\imath}\hat{\jmath}\hat{\sigma}}+\left[ X_{\hat{1}%
}\partial _{1},2\sum\limits_{l=2}^{n}u_{1l}\right] +\left[
\sum\limits_{k=2}^{n}X_{\hat{1}\hat{k}}\partial _{1}\partial _{k},2u_{1k}%
\right]
\end{eqnarray*}

From the formulae for the system with $n=2$ we have%
\begin{equation*}
\left[ X_{\left\{ k,l\right\} },-\Delta _{\left\{ k,l\right\} }\right] =-%
\left[ \partial _{k}\partial _{l},2u_{kl}\right] ,\qquad \text{and}\qquad %
\left[ \partial _{1}\partial _{k},2u_{1k}\right] =-\left[ \Theta _{\left\{
1,k\right\} },H_{\left\{ 1,k\right\} }\right]
\end{equation*}%
and from Lemma 2 we have that for any $\sigma ^{\prime }\in \mathfrak{S}%
\left( \hat{1};2t\right) ,$ $t\geq 2,$%
\begin{equation*}
\left[ X_{\sigma ^{\prime }},-\Delta _{\sigma ^{\prime }}\right]
=-2\sum\limits_{\substack{ k<l  \\ k,l\in \sigma ^{\prime }}}\left[
X_{\sigma ^{\prime }\backslash \left\{ k,l\right\} }\partial _{k}\partial
_{l},u_{kl}\right]
\end{equation*}%
and%
\begin{equation*}
\sum\limits_{t=1}^{\left[ \frac{n-3}{2}\right] }\sum\nolimits_{j=2}^{n}\sum%
\limits_{\sigma \in \mathfrak{S}\left( \sigma ^{\prime }\backslash \left\{
j\right\} ;2t\right) }\left[ X_{\sigma }\partial _{j},2\sum\limits_{l\in
\sigma }u_{jl}\right] \Theta _{\hat{\jmath}\hat{\sigma}}=0.
\end{equation*}%
Hence, $\left[ I,H\right] $ can be simplified to%
\begin{eqnarray*}
\left[ I,H\right] &=&\left[ \Theta ,H\right] +\sum\limits_{t=1}^{\left[ 
\frac{n-2}{2}\right] }\sum\limits_{\sigma \in \mathfrak{S}\left( \hat{1}%
;2t\right) }\left[ \Theta _{\hat{\sigma}},H_{\hat{\sigma}}\right] X_{\sigma
}+\sum\limits_{t=1}^{\left[ \frac{n-2}{2}\right] }\sum\limits_{\sigma \in 
\mathfrak{S}\left( \hat{1};2t\right) }\left[ \Theta _{\hat{\sigma}%
},2\sum\limits_{j\in \sigma }u_{1j}\right] X_{\sigma } \\
&&+\sum\nolimits_{j=2}^{n}\left[ \Theta _{\hat{\jmath}}\partial
_{j},2\sum\limits_{\substack{ l=1  \\ l\neq j}}^{n}u_{jl}\right]
+\sum\nolimits_{j=2}^{n}\sum\limits_{t=1}^{\left[ \frac{n-3}{2}\right]
}\sum\limits_{\sigma \in \mathfrak{S}\left( \hat{1}\hat{\jmath};2t\right) }%
\left[ \Theta _{\hat{\jmath}\hat{\sigma}}\partial _{j},2\sum\limits 
_{\substack{ l=1  \\ l\notin j\cup \sigma }}^{n}u_{jl}\right] X_{\sigma } \\
&&+\left[ X_{\hat{1}}\partial _{1},2\sum\limits_{l=2}^{n}u_{1l}\right]
-\sum\limits_{k=2}^{n}\left[ \Theta _{\left\{ 1,k\right\} },H_{\left\{
1,k\right\} }\right] X_{\hat{1}\hat{k}}.
\end{eqnarray*}%
This expression is equal zero. To verify this we should use relation (\ref%
{X_no_def_formula}), apply Lemma 6, which gives us%
\begin{equation*}
\left[ \Theta ,H\right] =-\sum\limits_{j=2}^{n}\left[ \Theta _{\hat{\jmath}%
}\partial _{j},2\sum\limits_{\substack{ l=1  \\ l\neq j}}^{n}u_{jl}\right]
-\sum\limits_{2\leq k<l\leq n}\left[ \Theta _{\hat{k}\hat{l}},2\left(
u_{1k}+u_{1l}\right) \right] X_{\left\{ k,l\right\} }
\end{equation*}%
and%
\begin{equation*}
\left[ \Theta _{\hat{\sigma}},H_{\hat{\sigma}}\right] =-\sum\limits 
_{\substack{ j=2  \\ j\notin \sigma }}^{n}\left[ \Theta _{\hat{\sigma}\hat{%
\jmath}}\partial _{j},2\sum\limits_{\substack{ l=1  \\ l\neq j,\sigma }}%
^{n}u_{jl}\right] -\sum\limits_{\substack{ 2\leq k<l\leq n  \\ k,l\notin
\sigma }}\left[ \Theta _{\hat{\sigma}\hat{k}\hat{l}},2\left(
u_{1k}+u_{1l}\right) \right] X_{\left\{ k,l\right\} },
\end{equation*}%
and equality%
\begin{equation}
\left[ \Theta _{\left\{ 1,k,l\right\} },H_{\left\{ 1,k,l\right\} }\right] =-%
\left[ \partial _{1},2\left( u_{1k}+u_{1l}\right) \right] X_{\left\{
k,l\right\} }-\left[ \Theta _{\left\{ 1,l\right\} }\partial _{k},2\left(
u_{1k}+u_{kl}\right) \right] -\left[ \Theta _{\left\{ 1,k\right\} }\partial
_{l},2\left( u_{1l}+u_{kl}\right) \right] ,  \label{equality_theta_n=3}
\end{equation}%
The last equality can be obtained by direct calculation or using formulae
for the three-particle case. Once all these substitutions are made it can be
easily seen that all terms in (\ref{tails}) are canceled. This completes the
proof of Theorem~1.

\section{Concluding remarks.}

A more general class of the deformed CM operators related to any Lie
superalgebra has been recently introduced in \cite{SV2004} by Sergeev and
Veselov, who found also recurrent formulae for the quantum integrals in
trigonometric case for the classical series. In the elliptic case the
integrability of these systems is an open problem. Within this approach the
operator (\ref{operator}) considered in our paper corresponds to the Lie
superalgebra $sl(n-1,1).$ The author hopes that the technique developed in
this paper can be applied to more general deformed elliptic CM operators
related to Lie superalgebra $sl(n,p).$

In this relation it is worth mentioning that as one can see from our
formulae (\ref{D_formulae}) the parameters $m=1/l,l=1,2,3\dots $ play a
special role in the theory of the deformed CM operators. It is interesting
to compare this with the results of the papers \cite{SV2004, SV}, where
exactly these particular values of the parameter arise in relation with the
strata in the discriminant variety consisting of polynomials having a root
of multiplicity $l.$

\section{Acknowledgement}

I am grateful to O. A. Chalykh and A. P. Veselov for attracting my attention
to this problem and useful discussions.


\begin{thebibliography}{99}
\bibitem{CFV1996} Chalykh O.A., Feigin M.V. and Veselov A.P., New integrable
deformation of quantum Calogero--Moser problem, Usp. Mat. Nauk 51(3),
185-186 (1996).

\bibitem{CFV1998} Chalykh O.A., Feigin M.V. and Veselov A.P., New integrable
generalizations of Calogero--Moser quantum problem, Journal of Math. Physics
39(2), 5341-5355 (1998).

\bibitem{KP2001} Khodarinova L.A.\ and Prikhodsky I.A., On algebraic
integrability of the deformed elliptic Calogero-Moser problem, J. of Nonlin.
Math. Phys. 8(1), 50-53 (2001).

\bibitem{CEO2003} Chalykh O.A., Etingof P. and Oblomkov A., Generalized Lam%
\'{e} operators, Commun. Math. Phys. 239, 115--153 (2003).

\bibitem{CMR1975} Calogero F., Marchioro C. and Ragnisco O., Exact solution
of the classical and quantal one-dimensional many-body problems with the
two-body potential $V_{a}(x)=g^{2}a^{2}/sh^{2}ax$, Lett.Nuovo Cim. 13(10),
383--387 (1975).

\bibitem{SK1975} Sawada K., Kotera T. Integrability and solution for the
one-dimensional N-particle system with inversely quadratic pair potentials.
J. Phys. Soc. Japan 39, 1614-1618 (1975).

\bibitem{OP1977} Olshanetsky M.A. and Perelomov A.M., Quantum completely
integrable systems connected with semi-simple Lie algebras. Lett. Math.
Phys. 2, 7-13 (1977).

\bibitem{OP1983} Olshanetsky M.A. and Perelomov A.M., Quantum integrable
systems related to Lie algebras. Phys. Reports 94, 313-404 (1983).

\bibitem{OOS1994} Ochiai H., Oshima T. and Sekiguchi H., Commuting families
of symmetrical differential-operators, Proc. of the Japan Academy, Ser. A -
Math. Sciences 70(2), 62-66 (1994).

\bibitem{OS1995} Oshima, T. and Sekiguchi H., Commuting families of
differential operators invariant under the action of a Weyl group, J. Math.
Sci. Univ. Tokyo 2(1), 1--75 (1995)

\bibitem{O1998} T. Oshima, Completely integrable systems with a symmetry in
coordinates, Asian J. Math 2, 935-955 (1998).

\bibitem{WW1927} Whittaker E.T. and Watson G.N., A course of modern
analysis, 4th edition, Cambridge University Press (1927).

\bibitem{Katz} Katz N. M., The congruences of Clausen-von Staudt and Kummer
for Bernoulli-Hurwitz numbers, Math. Ann. 216, 1--4 (1975).

\bibitem{K1998} Khodarinova L.A., On quantum elliptic Calogero-Moser
problem, Vestnik Mosc. Univ., Ser. Math. and Mech. 53(5), 16-19 (1998).

\bibitem{B1971} Berezin F.A., Laplace operators on semisimple Lie groups.
Proc. Moscow Math. Soc. 6, 371--463 (1971) (Russian)

\bibitem{SV2004} Sergeev A.N. and Veselov A.P., Deformed quantum
Calogero-Moser problems and Lie superalgebras, Commun. Math. Phys. 245,
249--278 (2004).

\bibitem{SV} Sergeev A.N. and Veselov A.P., Generalized discriminants,
deformed quantum Calogero-Moser systems and Jack polynomials,
math-ph/0307036.
\end{thebibliography}
\end{document}